\documentclass[twocolumn]{aastex631}
\usepackage{soul}
\usepackage{hyperref}

\newcommand\ha{H$\alpha$}
\newcommand\leg{{\em Legacy Survey}}
\newcommand\ngc{{NGC~5364}}
\newcommand{\mstarim}{$\rm \Sigma_\star$}
\newcommand{\sfrim}{$\rm \Sigma_{SFR}$}

\begin{document}

\title{Virgo Filaments V: Disrupting the Baryon Cycle in the NGC~5364 Galaxy Group}

\author[0000-0001-8518-4862]{Rose A. Finn}
\affiliation{Department of Physics and Astronomy, Siena College, 515 Loudon Road, Loudonville, NY 12211, USA}
\email{rfinn@siena.edu}

\author[0000-0001-5851-1856]{Gregory Rudnick}
\affiliation{University of Kansas, Department of Physics and Astronomy, 1251 Wescoe Hall Drive, Room 1082, Lawrence, KS 66049, USA}

\author{Pascale Jablonka}
\affiliation{Institute of Physics, Laboratory of Astrophysics, Ecole Polytechnique Fédérale de Lausanne (EPFL), Observatoire de Sauverny, CH-1290 Versoix, Switzerland}
\author{Mpati Ramatsoku}
\affiliation {INAF-Cagliari Astronomical Observatory, Via della Scienza 5, I-09047 Selargius (CA), Italy}
\affiliation{Department of Physics and Electronics, Rhodes University, P.O. Box 94, Makhanda, 6140, South Africa}
\affiliation{South African Radio Astronomy Observatory, 2 Fir Street, Black River Park, Observatory, Cape Town, 7405, South Africa}

\author[0000-0002-0905-342X]{Gautam Nagaraj}
\affiliation{Institute of Physics, Laboratory of Astrophysics, Ecole Polytechnique Fédérale de Lausanne (EPFL), Observatoire de Sauverny, CH-1290 Versoix, Switzerland}

\author[0000-0003-0980-1499]{Benedetta Vulcani}
\affiliation{INAF - Osservatorio astronomico di Padova, Vicolo Osservatorio 5, I-35122 Padova, Italy}

\author[0000-0002-3144-2501]{Rebecca A. Koopmann}
\affiliation{Department of Physics \& Astronomy, Union College, Schenectady, NY, 12308, USA}

\author{Matteo Fossati}
\affiliation{Dipartimento di Fisica G. Occhialini, Universit\`a degli Studi di Milano-Bicocca, Piazza della Scienza 3, I-20126 Milano, Italy}

\author[0000-0002-8122-3032]{James Agostino}
\affiliation{Ritter Astrophysical Research Center and Department of Physics \& Astronomy, University of Toledo, Toledo, OH 43606, USA}

\author{Yannick Bah\'{e}}
\affiliation{Institute of Physics, Laboratory of Astrophysics, Ecole Polytechnique Fédérale de Lausanne (EPFL), Observatoire de Sauverny, CH-1290 Versoix, Switzerland}

\author{Santiago Garcia-Burillo}
\affiliation{Observatorio Astronomico Nacional, Madrid, Spain}

\author[0000-0001-6831-0687]{Gianluca Castignani}
\affiliation{INAF - Osservatorio  di  Astrofisica  e  Scienza  dello  Spazio di  Bologna,  via  Gobetti  93/3,  I-40129,  Bologna,  Italy}

\author[0000-0003-2658-7893]{Francoise Combes}
\affiliation{Observatoire de Paris, LERMA, Coll\`ege de France, CNRS, PSL University, Sorbonne University, 75014, Paris, France}

\author[0009-0005-0303-0330]{Kim Conger}
\affiliation{University of Kansas, Department of Physics and Astronomy, 1251 Wescoe Hall Drive, Room 1082, Lawrence, KS 66049, USA}

\author[0000-0002-6220-9104]{Gabriella De Lucia}
\affiliation{INAF - Astronomical Observatory of Trieste, via G.B. Tiepolo 11, I-34143 Trieste, Italy}

\author{Vandana Desai}
\affiliation{IRSA, California Institute of Technology, MS 220-6, Pasadena, CA 91125, USA }

\author{John Moustakas}
\affiliation{Department of Physics and Astronomy, Siena College, 515 Loudon Rd, Loudonville, NY 12211, USA}

\author{Dara Norman}
\affiliation{National Optical Astronomy Observatory, 950N Cherry Avenue, Tucson, AZ 85750}

\author{Damien Sperone-Longin}
\affiliation{Institute of Physics, Laboratory of Astrophysics, Ecole Polytechnique Fédérale de Lausanne (EPFL), Observatoire de Sauverny, CH-1290 Versoix, Switzerland}

\author{Melinda Townsend}
\affiliation{Department of Physics and Astronomy, Siena College, 515 Loudon Rd, Loudonville, NY 12211, USA}

\author[0000-0003-3864-068X]{Lizhi Xie}
\affiliation{Tianjin Astrophysics Center, Tianjin Normal University, Binshuixidao 393, 
300387 Tianjin, People's Republic of China\\}

\author[0009-0001-1809-4821]{Daria Zakharova}
\affiliation{Dipartimento di Fisica e Astronomia Galileo Galilei, Universit\`a degli studi di Padova, Vicolo dell’Osservatorio, 3, I-35122 Padova, Italy
}
\affiliation{INAF - Osservatorio astronomico di Padova, Vicolo Osservatorio 5, I-35122 Padova, Italy}

\author[0000-0002-5177-727X]{Dennis Zaritsky}
\affiliation{Steward Observatory, University of Arizona, 933 North Cherry Avenue, Tucson, AZ 85721-0065, USA\\}

\begin{abstract}

The Virgo Filament Survey (VFS) is a comprehensive study of galaxies that reside in the extended filamentary structures surrounding the Virgo Cluster, out to 12 virial radii.  The primary goal is to characterize all of the dominant baryonic components within galaxies and to understand whether and how they are affected by the filament environment.  A key constituent of VFS is a narrowband \ha \ imaging survey of over 600 galaxies, VFS-\ha. The \ha \ images reveal detailed, resolved maps of the ionized gas and massive star-formation. This imaging is particularly powerful as a probe of environmentally-induced quenching because different physical processes affect the spatial distribution of star formation in different ways.  In this paper, we present the first results from the VFS-\ha\ for the NGC~5364 group,  a low-mass ($\log_{10}(M_{dyn}/M_\odot) < 13)$ system located at the western edge of the Virgo~III filament.  We combine \ha \ imaging with resolved H~I observations from MeerKAT for eight group members. These galaxies exhibit peculiar morphologies, including strong distortions in the stars and the gas, truncated H~I and \ha\ disks, H~I tails, extraplanar \ha\ emission, and off-center \ha\ emission. These signatures are suggestive of environmental processing such as tidal interactions,  ram pressure stripping, and starvation. We quantify the role of ram pressure stripping expected in this group, and find that it can explain the cases of H~I tails and truncated \ha \ for all but one of the disk-dominated galaxies.  Our observations indicate that multiple physical mechanisms are disrupting the baryon cycle in these group galaxies.

\end{abstract}

\keywords{Galaxy Groups; Galaxy Quenching; Galaxy Environments; Large-scale structure of the universe; Cosmic web}

\section{Introduction} \label{sec:intro}

The physical properties of galaxies are strongly correlated with their environment; galaxies residing in high-density regions tend to host older stellar populations, have lower star formation rates (SFRs), and different morphologies than galaxies in the field \citep[e.g.,][]{Dressler1980,Blanton2009}.   The alteration of the gas contents and SFRs is invariably linked to the disruption of the baryon cycle, the exchange of gas between the interstellar medium (ISM) and intergalactic medium (IGM) and its conversion to stars within galaxies \citep[e.g.,][]{Tumlinson2017}.

Understanding the disruption of the baryon cycle requires mapping all the components of the gas as well as those of the stars, because the abundance and spatial distribution of different components respond differently to physical processes. 
For example, strong active galactic nuclei (AGN) and supernova feedback may quench star formation through ejection or heating of the dense gas and the reservoir of atomic gas without impacting the distribution of preexisting stars 
\citep[e.g.,][]{Springel2005,Croton2006,Dekel2006}.  Processes such as tidal interactions and mergers can affect the distribution of both stars and multiple phases of the gas \citep{Toomre1972,Mihos1993, Semczuk2020}, whereas pressure-driven interactions, such as starvation \citep[e.g.,][]{Larson1980}, remove the hot gas halo and decouple galaxies from their filamentary connections to the cosmic web while leaving the stars untouched.  Ram pressure stripping \citep[e.g.,][]{Gunn1972}, on the other hand, can remove diffuse neutral gas \citep{Kenney2004,Chung2007,Chung2009,Jaffe2015,Ramatsoku2019}, ionized gas \citep{Koopmann2004,Poggianti2019}, or even potentially dense molecular gas \citep{Boselli2014,Scott2015,Lee2017} from the galaxy disk.  However, ram pressure can also compress H~I and {facilitate its conversion} into H$_2$ \citep{Moretti2023}.
Studying the relative spatial distribution and amounts of different gas phases and of the stars can therefore help identify the dominant processes that deplete gas in galaxies.

{While historically attention has focused on how the baryon cycle is disrupted in clusters, groups, and the field, more recently the community has turned its attention to the filamentary network that feeds clusters and groups \citep[e.g.,][]{Cybulski2014, Just2019, Sarron2019,  Lee2021, Castignani2022a, Salerno2022}. }
To further understand how galaxies are altered over the full dynamic range of local environmental densities, we are undertaking a multiwavelength study of galaxies in and around the cosmic filaments surrounding the Virgo cluster, extending up to $\sim$12 virial radii in projection from one of the best-studied regions of high density in the local Universe.  {The first set of papers from the VFS found results that demonstrate that filaments are sites in which galaxies have their gas content and morphology changed \citep{Castignani2022a,Castignani2022b}.  However, \citet{Zakharova2024} found in another VFS paper that gas depletion is being driven primarily by groups embedded within filaments.  Additionally, \citet{Conger2025} find that late-type galaxies in the densest VFS environments have smaller dust disks than in low-density environments. There is also preliminary evidence of smaller star-forming disks even in intermediate-density environments, i.e. filaments and groups.  Given these results, it is still uncertain what relative role filaments, and the groups that are embedded within them, play in transforming galaxies.}

\begin{figure*}
    \centering
    \includegraphics[width=0.95\textwidth]{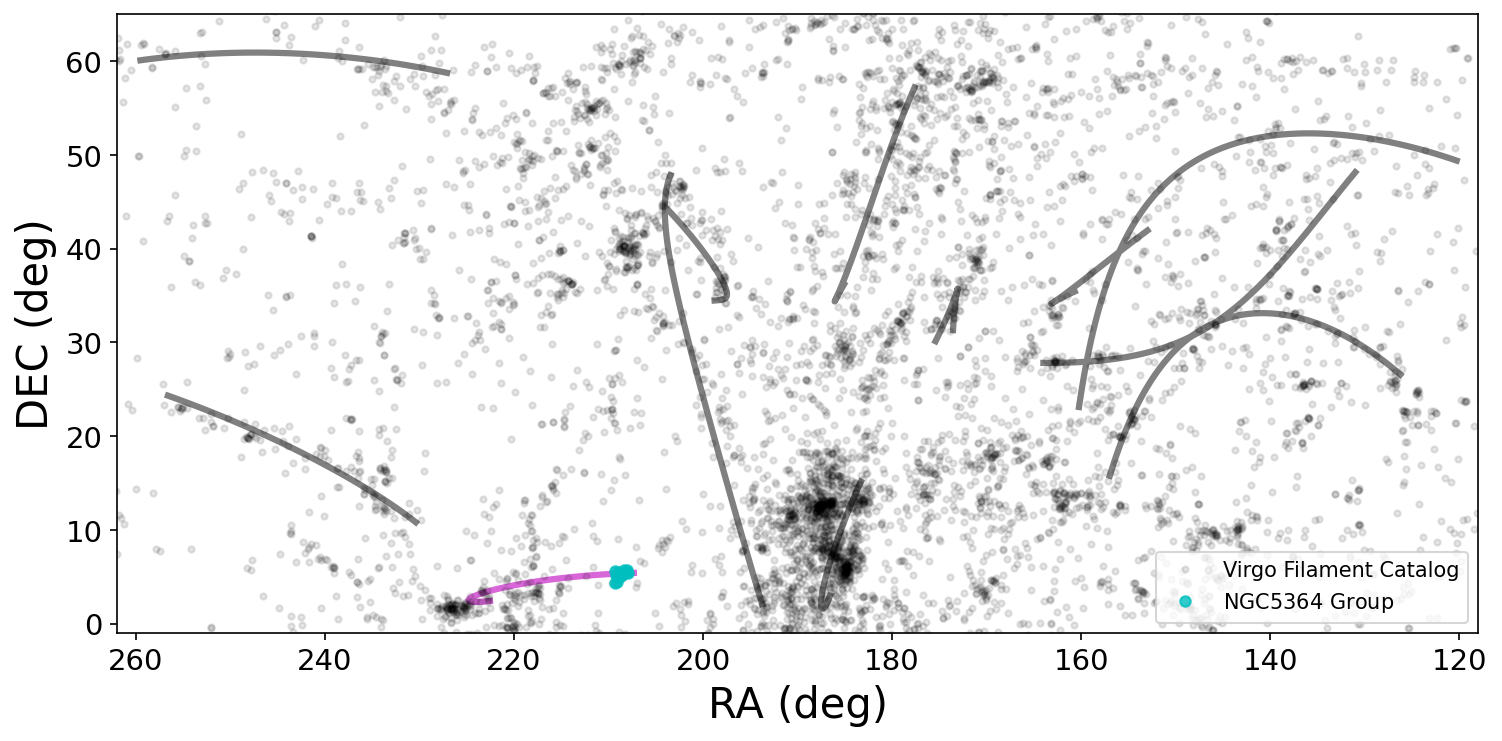}
    \caption{Location of the NGC~5364 Group on the plane of the sky relative to the Virgo Cluster and surrounding large-scale structure covered by the {\it Virgo Filament Survey}.  The magenta line shows the Virgo~III filament spine, and the gray lines show the spines of the other filaments identified by \citet{Castignani2022b}.  The NGC~5364 group galaxies are shown with the cyan circles.  The group is located at the western end of the Virgo~III Filament and approximately 6~Mpc from the center of Virgo. The Virgo cluster is the large overdensity near ($\rm R.A.=185^\circ$, $\rm decl.=10^\circ$).}
    \label{fig:lss}
\end{figure*}

A crucial next step in understanding how the gas content of galaxies is modified in the cosmic web is to characterize the detailed spatial distribution of star formation and to link it with the properties of the multiphase ISM.  To that end, we have completed a large \ha \ narrowband imaging survey of over 600 galaxies in the extended volume around Virgo, including all the galaxies with CO measurements \textbf{from} \cite{Castignani2022a}. 
In this paper, we present the first results from this survey, the Virgo Filament Survey-\ha \ (VFS-\ha),  focusing solely on the NGC~5364 group.  This  group was targeted in our \ha \ survey because it is located in the Virgo III filament \citep{Kim2016,Castignani2022b}, and we show the location of the NGC~5364 Group relative to the Virgo Cluster and surrounding large-scale structure in Figure \ref{fig:lss}.  We have also obtained spatially resolved H~I imaging from MeerKAT (Ramatsoku et al. in prep.).  We will present the full VFS-\ha \ survey results in subsequent papers.  In this paper we leverage our multiband data for the NGC~5364 group to illustrate the variety of ways in which the baryon cycle can be disrupted in even the lowest-mass galaxy groups.

In Section~\ref{sec:ngcgroup} we describe the group and its place within the larger context of the extended Virgo volume.  In Section~\ref{sec:data} we describe the \ha\ data and resolved maps of the SFR, stellar mass, and H~I. 
 We present our results in Section~\ref{sec:results} and discuss them in Section~\ref{sec:discuss}.  We summarize in Section~\ref{sec:summary}.  Throughout this paper we adopt a $\Lambda$CDM cosmology with $\Omega_m=0.3,~\Omega_\Lambda=0.7$ and $H_0 = 70~{\rm km~s^{-1}~Mpc^{-1}}$.  We assume a \citet{Chabrier2003} Initial Mass Function throughout.

\section{The NGC 5364 Group}
\label{sec:ngcgroup}

The NGC~5364 group is listed in several catalogs \citep[e.g.,][]{Kourkchi2017,Lim2017,Tempel2017}. 
Following \citet{Castignani2022b}, we adopt the group membership and properties from \citet{Kourkchi2017}. 
\citet{Kourkchi2017} identify 17 group members, 
and the \ha \ image covers eight of these, including the five most massive galaxies within the group (see Table \ref{tab:sample} and the layout in Fig. \ref{fig:INT_footprint}). 
{In Table \ref{tab:sample}, we include the Hubble T-type from Hyperleda \citep{Makarov2014} when available, in column 3.}
{Stellar masses and SFRs are derived from spectral energy distribution (SED) fitting using MAGPHYS \citep{daCunha2008} and custom near-ultraviolet (NUV) through IR photometry. 
 The details of the SED fitting are presented in \citet{Conger2025}, and the resulting values are listed in Table~\ref{tab:sample}.} The group galaxies outside the field of view (FOV) all have $\log_{10}(M_\star/M_\sun) < 8.9$.
 We calculate the stellar mass-weighted center using all 17 members (R.A. = 208.9912 deg, decl. = 5.11883 deg) and show the location with the cyan X in Figure \ref{fig:legacy}.   The center of the group is nicely placed within the \ha \ image.

According to \citet{Kourkchi2017}, the NGC~5346 Group 
has a line-of-sight velocity dispersion of $\sigma_V=155$~km/s, and we reach a similar estimate of the velocity dispersion using the biweight scale of the 17 group members (164~km/s). 
The dynamical mass given in \citet{Kourkchi2017} is $\rm \log_{10}(M_{halo}/M_\sun) = 12.7$.  {Thus this is an extremely low-mass group. The biweight location of the flow-corrected recession velocities \citep{Castignani2022b} is 1425~km~s$^{-1}$, which corresponds to a distance of $\sim$20~Mpc.}

\begin{table*}
\begin{center}
\caption{NGC~5364 Group Members within the \ha \ Footprint \label{tab:sample}}
\begin{tabular}{cccccccccc}
\hline \hline
VFID & NED Name & $T^{a}$ & $v_r^{b}$ & Virgo $d_{3d}^{c}$ & $1/T(z)^{d}$ & $\log(M_\star)^e$ & $\log(SFR)^f$ & $\log(sSFR)^g$ & H2 def$^h$ \\
 &  &  & $\mathrm{km\,s^{-1}}$ & $\mathrm{Mpc}$ &  & $\mathrm{M_{\odot}}$ & $\mathrm{M_{\odot}\,yr^{-1}}$ & $\mathrm{yr^{-1}}$ &  \\
\hline
VFID5842 & NGC 5356 & 3.8 & 1370 & 9.7 & 1.07 & 9.96 & -0.96 & -10.85 & -0.29 \\
VFID5844 & SDSS J135621.31+051944.2 & -2.9 & 1402 & 8.9 & 1.08 & 7.79 & -4.43 & -12.22 & \nodata \\
VFID5851 & NGC 5363 & 0.1 & 1134 & 6.3 & 1.04 & 10.84 & -1.83 & -12.60 & -0.19 \\
VFID5855 & NGC 5348 & 3.8 & 1450 & 5.5 & 1.08 & 9.52 & -0.74 & -10.07 & 0.19 \\
VFID5859 & WISEA J135504.45+051121.7 & 10.0 & 1405 & 4.0 & 1.08 & 7.68 & -1.62 & -9.03 & \nodata \\
VFID5879 & SDSS J135502.70+050525.1 & -5.0 & 1397 & 8.8 & 1.07 & 8.40 & -4.60 & -12.90 & \nodata \\
VFID5889 & NGC 5364 & 4.0 & 1240 & 5.1 & 1.06 & 10.42 & -0.37 & -10.30 & 1.06 \\
VFID5892 & NGC 5360 & 0.1 & 1175 & 7.9 & 1.05 & 9.09 & -1.57 & -10.56 & 0.51 \\
\hline
\end{tabular}
\end{center}
\tablecomments{$^a$ Hubble T-type from Hyperleda \citep{Makarov2014}. \\
$^b$ Heliocentric recession velocity. \\
$^d$ Flow-corrected, 3D distance from the center of the Virgo Cluster from \citet{Castignani2022a} \\
$^d$ \ha \ flux correction, equal to the inverse of the filter transmission at the redshift of galaxy. \\
$^e$ Stellar mass from MAGPHYS SED fitting presented in \citet{Conger2025}. \\
$^f$ SFR from MAGPHYS SED fitting presented in \citet{Conger2025}. \\
$^g$ sSFR from MAGPHYS SED fitting presented in \citet{Conger2025}. \\
$^h$ $H_2$ deficiency from \citet{Castignani2022a}.\\
}
\end{table*}

\section{Data}
\label{sec:data}
\subsection{\ha \ Observations}
\begin{figure}
    \centering
    \includegraphics[width=0.5\textwidth]{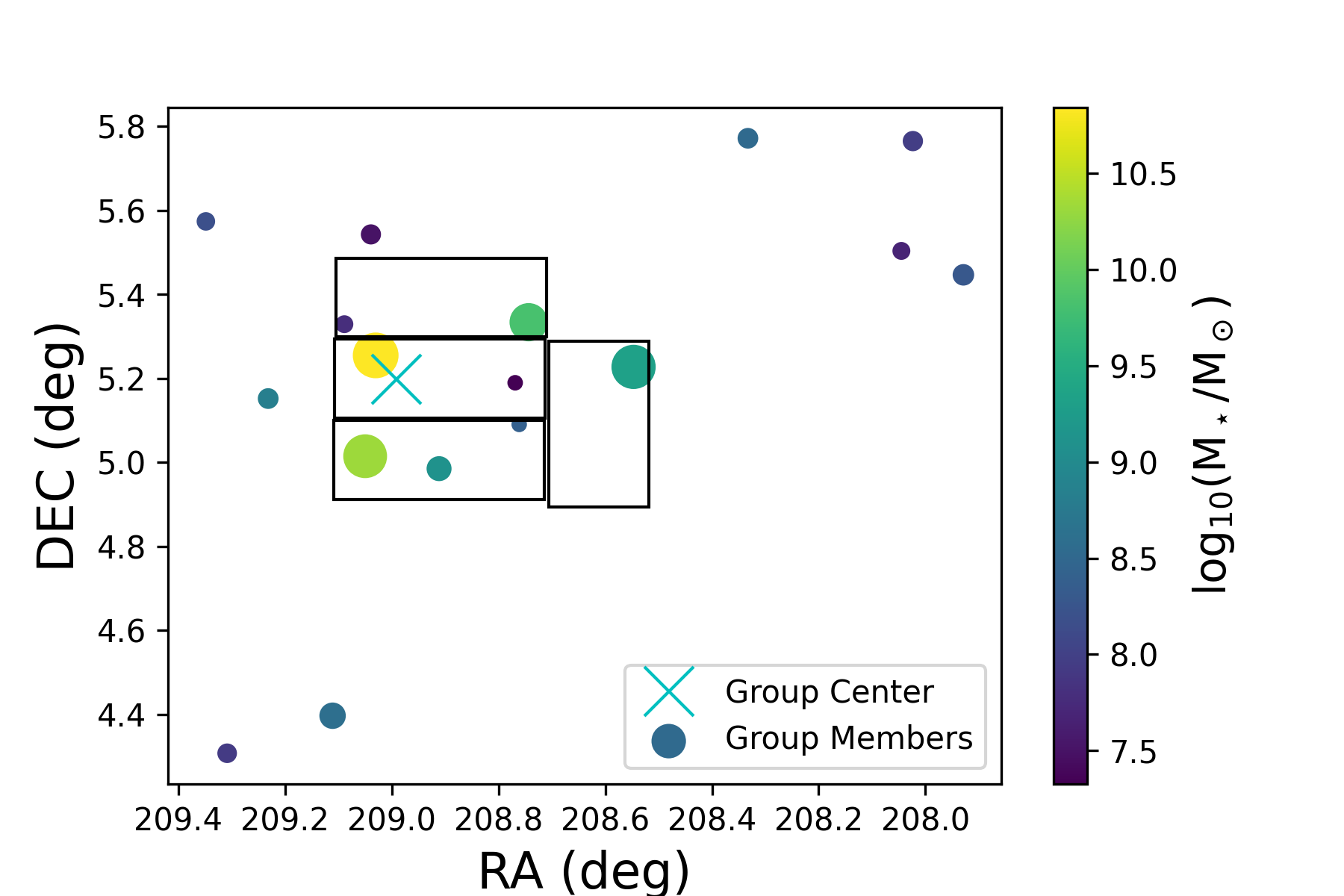}
    \caption{Decl. vs. R.A. for the 17 NGC~5364 group members.  Circles are color coded by stellar mass, and the size of the circles scales with the isophotal size of the galaxies.  The black rectangles show the approximate location of the four detectors of the Wide Field Camera.  The eight group members within the \ha \ field of view are the focus of this paper.  The cyan X shows the mass-weighted center of the group.}
    \label{fig:INT_footprint}
\end{figure}

As part of our larger Virgo Filament \ha\ survey, the NGC 5346 group was imaged 
with the Wide Field Camera (WFC) at the Isaac Newton Telescope on 2019 February 06.  The WFC is a mosaic camera containing four $2048\times4096$ CCDs, and the field of view is 34\arcmin$\times34$\arcmin.  Three of the CCDs are arranged horizontally with the long dimension running east-west, and the fourth CCD is positioned west of the first three and is oriented vertically, with the long dimension running north-south.  As a result, the CCDs do not cover the northwest corner of the field of view.  We show the WFC footprint with the NGC~5364 group members in Figure \ref{fig:INT_footprint}.  {The pixel scale of the WFC is 0.333\arcsec \ per pixel, which corresponds to a physical size of $\simeq$33~pc~pixel$^{-1}$ at the distance of the \ngc \ group.}

We use the Sloan Digital Sky Survey (SDSS) $r$-band filter to measure the stellar continuum and a narrowband filter to detect \ha \ emission.
The \ha \ filter has a central wavelength of 
6568\AA \ and a bandwidth of 95\AA, so it can detect \ha \ from galaxies with recession velocities in the range $-1942$~km~s$^{-1}~ < v_r < 2400$~km~s$^{-1}$.  
The 30 minutes \ha \ integration was comprised of ten 180~s exposures taken in a five-position dither pattern with approximately 30\arcsec \ offsets.  The total $r$-band exposure was five minutes, consisting of five individual 60~s exposures taken at the same dither positions as the \ha \ images.  

The data reduction follows standard procedures.  
We complete bias subtraction and flattening with sky flats using the GUI interface for THELI v3 \citep{Erben2005,Schirmer2013}, and we combine the bias and flat images for an entire run to maximize the signal-to-noise ratio.    
We create reference catalogs using Source Extractor \citep{Bertin1996} and solve for the astrometric calibration using Scamp \citep{Bertin2010}.  In Scamp, we used the GAIA EDR3 as the reference catalog and fit the geometric distortion in the WFC images with a fourth-order polynomial.  
When running Scamp, we set the
STABILITY\_TYPE=EXPOSURE, MOSAIC\_TYPE=FIX\_FOCALPLANE.  
The astrometric RMS error with respect to the GAIA EDR3 reference positions is significantly less than a pixel and is typically 0.05\arcsec \ or less.
We determine the AB photometric zero-point by comparing instrumental magnitudes with the reference $r$-band magnitude from PANSTARRS for both the \ha\ and $r$-band image.  Calibrating both of our images to the same PANSTARRS filter allows us to use the $r$-band image as our continuum source, as we describe in Section \ref{sec:SFRmap}.

\subsection{\ha \ Star Formation Maps}
\label{sec:SFRmap}

To create the continuum-subtracted \ha \ image, 
we follow the procedure described in detail in \citet{Boselli2018} \citep[see also, e.g.,][]{Kennicutt2008}. We use a scaled version of the $r$-band image to estimate the continuum in the \ha \ image, and to first order, the scale factor is determined from the difference in photometric zero-points in the \ha \ and $r$-band images.  The exact scale factor also depends on the color of the stellar continuum, so we adjust the scale factor by a term that scales with the $g-r$ color variations within the galaxy.  The $g-r$ color dependence is determined by integrating a large library of stellar spectra over the $r$ and \ha \ filters (see \citealt{Boselli2018} for details).  
We did not obtain our own $g$-band images, so we constructed a $g-r$ image by reprojecting the Dark Energy Spectroscopic Instrument (DESI) {\it Legacy} Imaging Surveys (hereafter known as the {\it Legacy Surveys}) $g$ and $r$ images \citep{Dey2019} to match the field of view (FOV) and pixel scale of the \ha \ image. The surface brightness corresponding to the standard deviation in the sky noise in the continuum-subtracted \ha \ image is $\rm 2.9 \times 10^{-17}~erg~s^{-1}~cm^{-2}~arcsec^{-2}$. 

We convert the continuum-subtracted \ha \ image to an SFR image using the calibration from \citet{Kennicutt2012}.  {The conversion assumes the IMF as described in \citet{Kroupa2003}, which includes a turnover below $\sim 1M_\sun$.  This IMF produces similar results to the Chabrier IMF \citep{Chabrier2003}, as demonstrated by \citet{Chomiuk2011}.}  We correct for the fact that the \ha \ emission is within the $r$-band filter, which means that the scaled $r$-band image overestimates the continuum \citep[e.g.,][]{Kennicutt2008}. An additional correction arises because the transmission across the \ha \ filter is not uniform.  We use the recession velocity of the galaxy and the filter trace for the \ha \ filter to determine the filter transmission at the observed wavelength of \ha\ and use this to correct for the variation in the transmission with wavelength.  {The filter correction is less than 10\%, and we list the correction applied to each galaxy in column 5 of Table \ref{tab:sample}.} 
Finally, we correct the \ha \ flux for Milky Way extinction, but not for internal extinction within the galaxy.
{We note that the uncertainty in the resulting SFR maps can be systematically larger in regions where the SFR is low. As discussed in detail in \citet{Kennicutt2012}, the correlation between \ha \ emission and the SFR can break down at low SFRs and on small spatial scales due to the low number density and short lifetimes of the massive O stars that drive the correlation.}

\subsection{Stellar Mass Maps}
We create stellar mass maps from optical imaging and a relationship that predicts the mass-to-light ratio from optical colors \citep[e.g.,][]{Bell2001, Bell2003,Zibetti2009,Roediger2015}.  In practice, we use the reprojected {\it Legacy} $g$ and $r$ images described in \S\ref{sec:SFRmap}.  We then convert the $g-r$ image to a mass-to-light ratio ($M/L$) image using the linear relation from \citet{Roediger2015}:
\begin{equation}
    \log_{10}(M_\star/L)_\lambda = m_r × (g-r) + b_r,
\end{equation}
where $m_r=1.629$ and $b_r=-0.792$.  This fit assumes a \citet{Chabrier2003} IMF.  The stellar mass image is created by first converting the $r$-band image from flux to luminosity, and then  multiplying by the $M/L$ image.  The approximate uncertainty associated with the resulting stellar masses is $\sim 0.2$dex \citep{Roediger2015}.
Finally, we convert the stellar mass per pixel to units of ${\rm \Sigma_\star} (M_\sun/\rm kpc^2)$ using the angular diameter distance that corresponds to the mass-weighted, average recession velocity of the group members.

\subsection{H~I Maps}

We obtained MeerKAT data on this group as part of a larger survey of the Virgo III filament (Ramatsoku et al. in prep).  A single MeerKAT pointing covers this group.  The data are primary beam corrected on a channel-by-channel basis and result in a 1$\sigma$ column density limit of $N_{H~I} = 3.4\times 10^{19} {\rm~cm^{-2}}$.  In this paper we focus on the H~I morphology and how it relates to the spatial distribution of \ha \ emission.  A more detailed and quantitative analysis of the H~I data will be presented in Ramatsoku et al. (2025, in preparation).

\section{Results}
\label{sec:results}
\subsection{Spatial Distribution of Star Formation}
\label{sec:results_spatdist}

\begin{figure*}
    \centering
    \includegraphics[width=\textwidth]{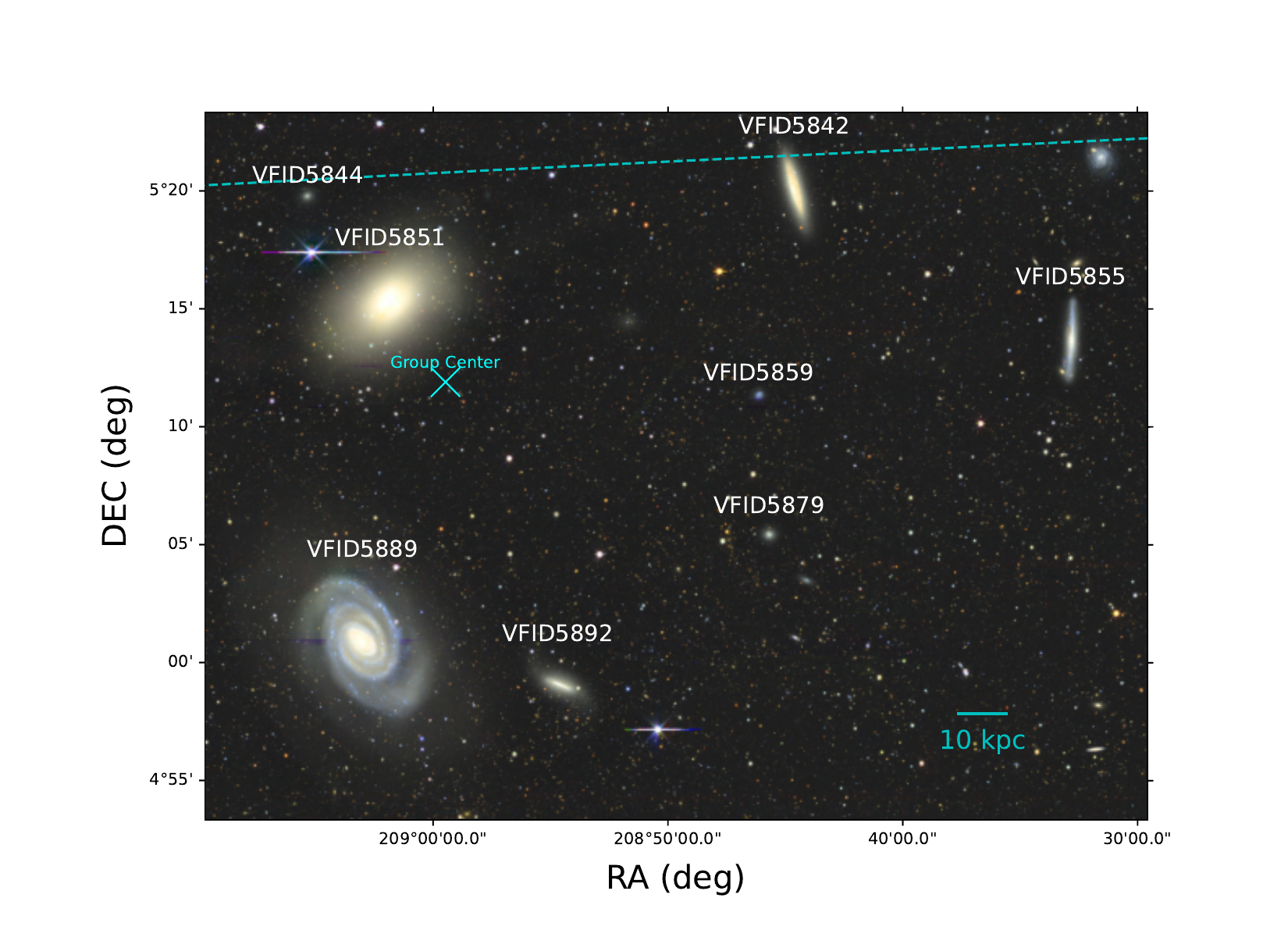}
    \caption{Legacy DR9 $grz$ image showing the location of group members.  The dashed cyan line shows the location of the Virgo III filament spine, projected in R.A. and decl.  Group members are labeled with their VFID \citep{Castignani2022b}.  
  }
    \label{fig:legacy}
\end{figure*}

We show the positions of group members that fall within the \ha \ FOV in Figure \ref{fig:legacy}.  
The dashed line shows the Virgo~III filament spine, projected onto the plane of the sky \citep{Castignani2022b}.  
Eight of the 17 group members are located within the footprint of the \ha \ image (see Fig. \ref{fig:INT_footprint}), and this paper focuses primarily on these galaxies.  

{In Figures \ref{fig:mstar_massive} through \ref{fig:mstar_dwarfs}, we show the relative distribution of the stellar mass and star formation for the eight galaxies within the \ha\ FOV.  We present the galaxies in order of decreasing stellar mass.  We discuss these results in the context of previous work in Section \ref{sec:processes}.}  

\subsubsection{The Two Most Massive Galaxies: VFID5851 and VFID5889}
\begin{figure*}
    \centering
    \includegraphics[width=\textwidth]{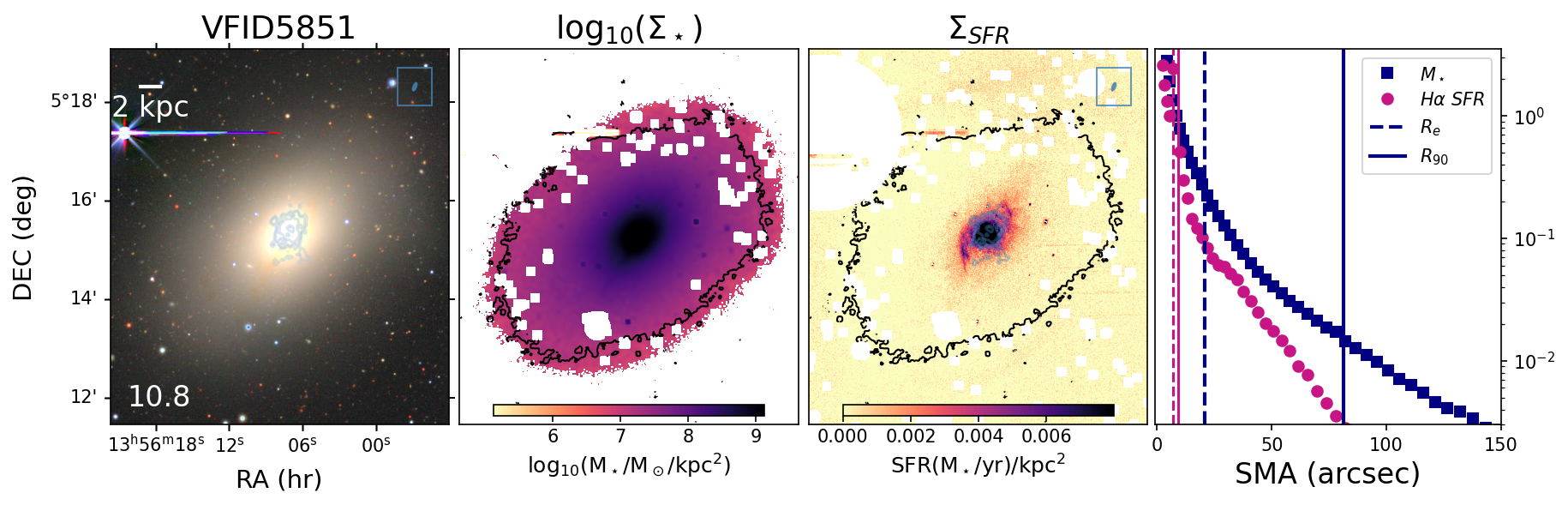}
    \includegraphics[width=\textwidth]{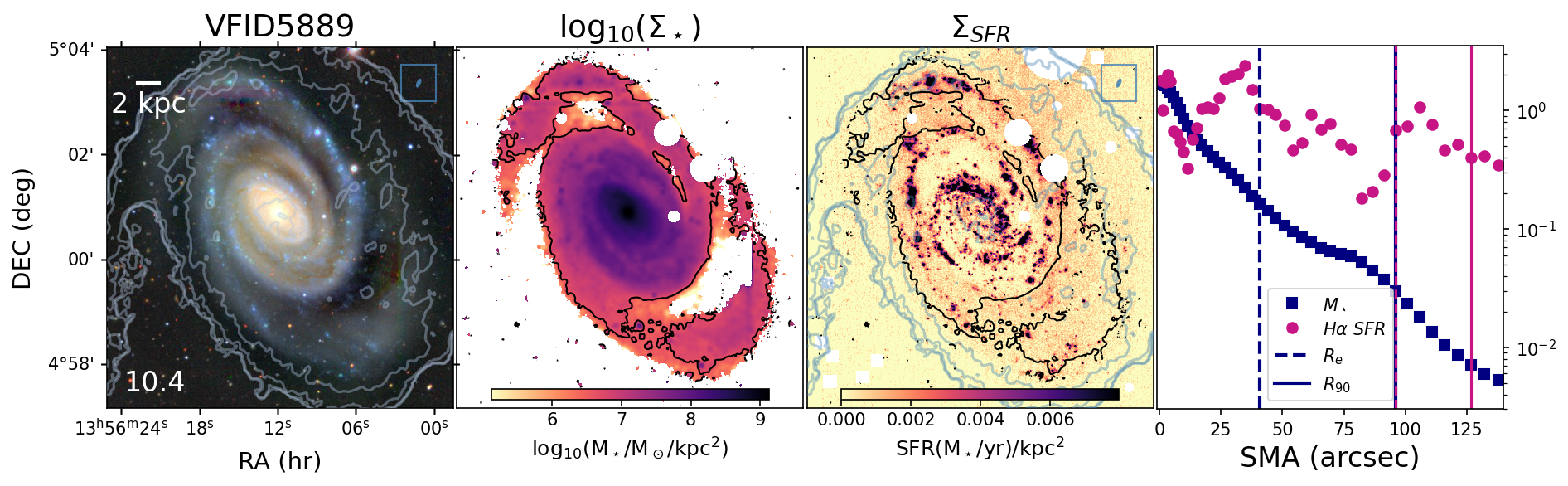}
    \caption{Results for the two most massive ($\log_{10}(M_\star/M_\odot) > 10$) galaxies in the group.  (Top) VFID5851, the most massive galaxy and central elliptical. The left panel shows the Legacy $grz$ color image, with $\log(M_\star/M_\sun)$ written in the lower left and a scale bar in the top left.  The second panel shows the stellar mass map, with the color bar showing $\log_{10}(M_\star/M_\sun/kpc^2)$.  White regions correspond to foreground stars or background galaxies that are masked for the image analysis.  The third panel shows  the \ha \ SFR map, with the color bar showing $\log_{10}(M_\star/M_\sun ~yr^{-1})/kpc^2$.  The fourth panel shows the radial profiles of the stellar mass and SFR, normalized to unity in the central 5\arcsec. The MeerKAT H~I contours are shown in gray in the first and third panels, and the beam size is shown with the blue ellipse in the top right of the first and third panels.  The black contours in the two middle panels show where  $\rm \log_{10}(M_\star/M_\sun/kpc^2) \approx 6.6$, and this is included to aid in comparing the spatial extent of the stellar mass and \ha \ SFR.   (Bottom) Same as above, showing results for VFID5889, a large spiral with normal \ha \ emission.  
    }
    \label{fig:mstar_massive}
\end{figure*}

In Figure \ref{fig:mstar_massive}, we show the results for the two most massive galaxies in the group.  Each row shows, from left to right, the \leg \ $grz$ color image with H~I contours from Ramatsoku et al. (2025, in preparation), the stellar mass surface density map, 
the \ha \ star-formation surface density map, and radial profiles of the stellar mass and \ha \ SFR.  The radial profiles are measured from elliptical aperture photometry, where we fix the ellipse geometry based on the stellar mass distribution and apply the same apertures when measuring the flux in the SFR image.  The dashed and solid vertical lines show the radii that enclose 50\% and 90\% of the flux, respectively, and these are measured from the elliptical aperture photometry.     

The top row of Figure \ref{fig:mstar_massive} shows the results for VFID5851, the most massive member of the group. The {\it Legacy} $grz$ image reveals its elliptical morphology, and the stellar mass distribution is fairly smooth.  Extended \ha \ emission is detected at the center, and    
the H~I contours reveal the presence of atomic gas in the same inner region.  
The radial profiles confirm that the \ha \ emission is more centrally concentrated than the stellar mass profile, and this is also reflected more quantitatively in the smaller values of $R_{50}$ and $R_{90}$ for the \ha-SFR profile.

The bottom row of Figure \ref{fig:mstar_massive} shows the results for VFID5889, the main spiral galaxy and the namesake of the group (NGC~5364).  In this case, the \ha\ emission follows the spiral arms, with no star formation in the interarm region.  The \ha \ emission is clumpy, which is likely due to HII regions within the galaxy. Comparison of the $\Sigma_\star$ and $\Sigma_{\rm SFR}$ images shows that the  \ha \ emission is similar in extent to the stellar mass.

\subsubsection{Intermediate-mass Spirals: VFID5842, VFID5855, VFID5892}
\begin{figure*}
    \centering
        \includegraphics[width=\textwidth]{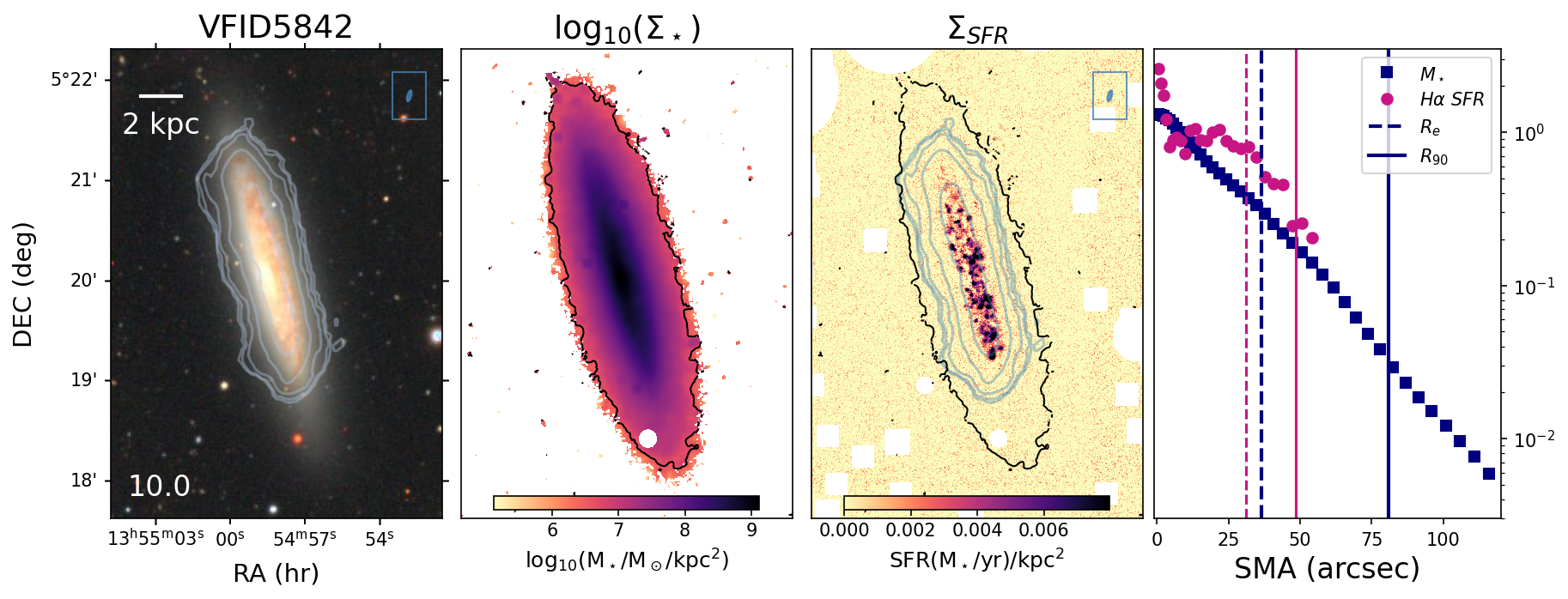}
 \includegraphics[width=\textwidth]{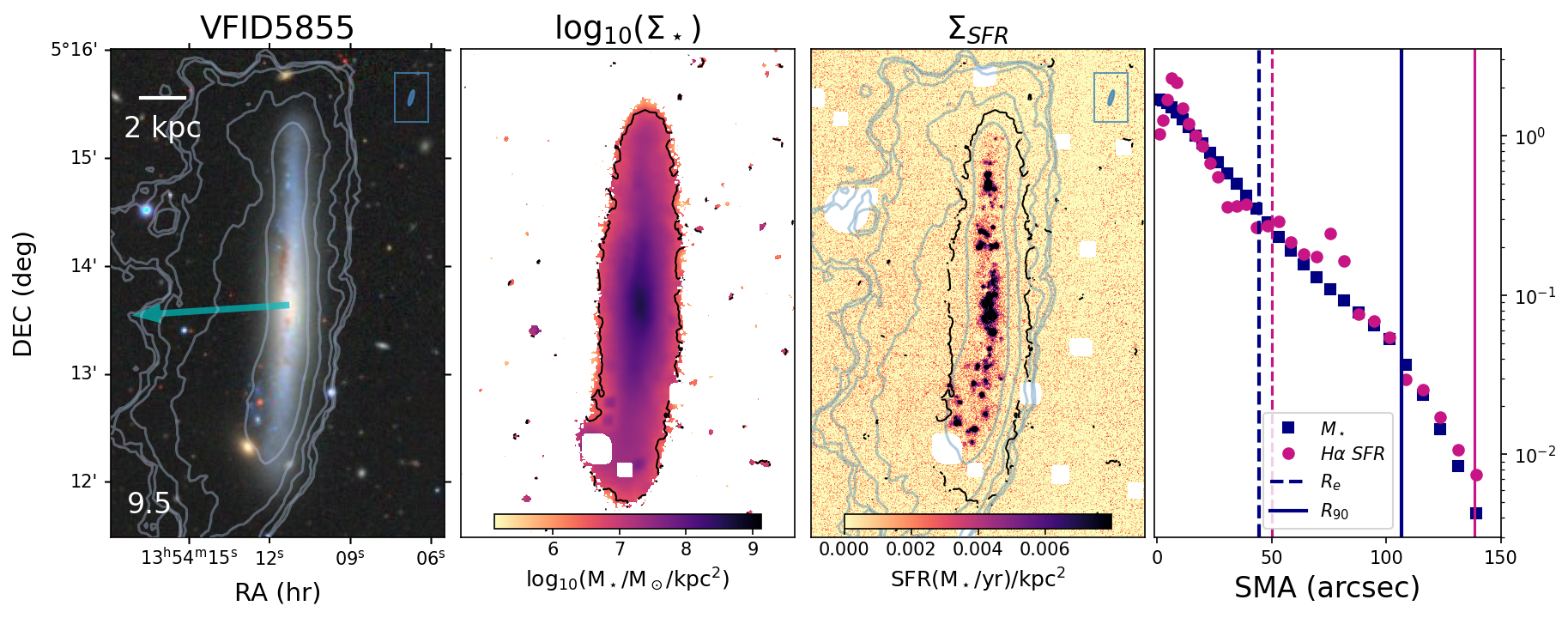}

    \includegraphics[width=\textwidth]{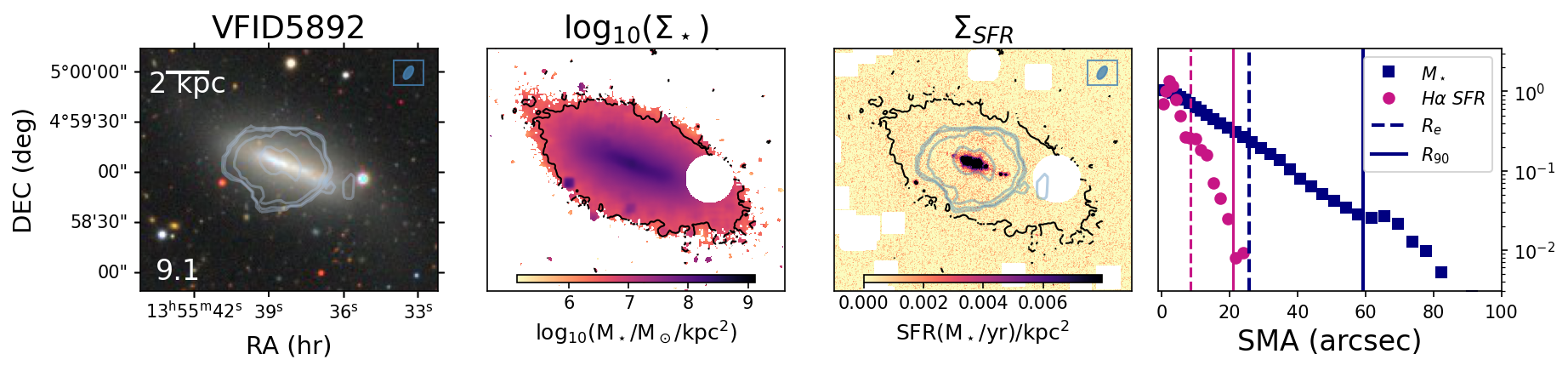}
         
    \caption{Results for the three $9 < \log_{10}(M_\star/M_\odot) \le 10$ group galaxies: VFID5842 (top), VFID5855 (middle), and VFID5892 (bottom). 
    The left panel shows the Legacy $grz$ color image, with $\log(M_\star/M_\sun)$ noted in the lower left and a scale bar in the top left.  The second panel shows the stellar mass map, with the color bar showing $\rm \log_{10}(M_\star/M_\sun/kpc^2)$.  The third panel shows  the \ha \ SFR map, with the color bar showing $\rm  \log_{10}(M_\star/M_\sun ~yr^{-1})/kpc^2$.  The fourth panel shows the radial profiles of the stellar mass and SFR, normalized to unity in the central 5\arcsec. The MeerKAT H~I contours are shown in gray in the first and third panels, and the beam size is shown with the blue ellipse in the top right of the first and third panels.  The black contours in the two middle panels show where  $\rm \log_{10}(M_\star/M_\sun/kpc^2) \approx 6.6$, and this is included to aid in comparing the spatial extent of the stellar mass and \ha \ SFR.   
    }
    \label{fig:mstar_spirals}
\end{figure*}

In Figure \ref{fig:mstar_spirals}, we show the results for three disk galaxies with $9 < \log_{10}(M_\star/M_\sun) \le 10 $, with the galaxies ordered from top to bottom by decreasing stellar mass.  
The top row of Figure \ref{fig:mstar_spirals} shows VFID5842, a highly inclined spiral.  Comparison of the \mstarim \ and \sfrim \ images shows that the \ha \ emission is clumpy and truncated relative to the stellar component.
The radial profiles in the right panel show that the \ha \ emission drops abruptly at a radius of approximately 60\arcsec \ or 7.7~kpc.   Interestingly, $R_{50}$ is similar for the stellar and SFR profiles, but $R_{90}$ is significantly smaller for the SFR profile.  This indicates that the outer part of the star-forming disk is being most affected.  The H~I contours are truncated relative to the stellar mass but are slightly larger in extent than the \ha \ emission.

The middle row in Figure \ref{fig:mstar_spirals} shows the results for VFID5855, an edge-on spiral.  While the \ha \ emission is clumpy, the stellar mass and SFR maps show similar radial extent, and the radial profiles show similar shapes.  The $\Sigma_{\rm SFR}$ image shows that the \ha \ emission is above the plane of the disk on the southern side of the galaxy.  Using custom apertures, we estimate that $\approx$15$\%$ of the total \ha \ emission is above the plane and that this extraplanar gas is clumpy (Nagaraj in prep). Interestingly, the H~I emission is highly asymmetric and extends beyond the disk on the same side as the extraplanar \ha \ emission.  The cyan arrow shows the direction to the group center, and the H~I and \ha \ extensions are aligned in this direction.  Note that the filament spine is north of the galaxy and runs perpendicular to the disk on the plane of the sky.  
 
 In the bottom panel of Figure \ref{fig:mstar_spirals}, we show VFID5892.  
As with VFID5842, comparison of the stellar mass and SFR images shows that the star formation is truncated relative to the stellar disk.
The radial profiles show that the \ha \ emission is more concentrated, and in this case, both $R_{50}$ and $R_{90}$ are significantly smaller for the SFR profile.
The H~I emission is also truncated relative to the stellar disk and shows similar extent to the \ha \ emission.

\begin{figure*}
    \centering  
         \includegraphics[width=\textwidth]{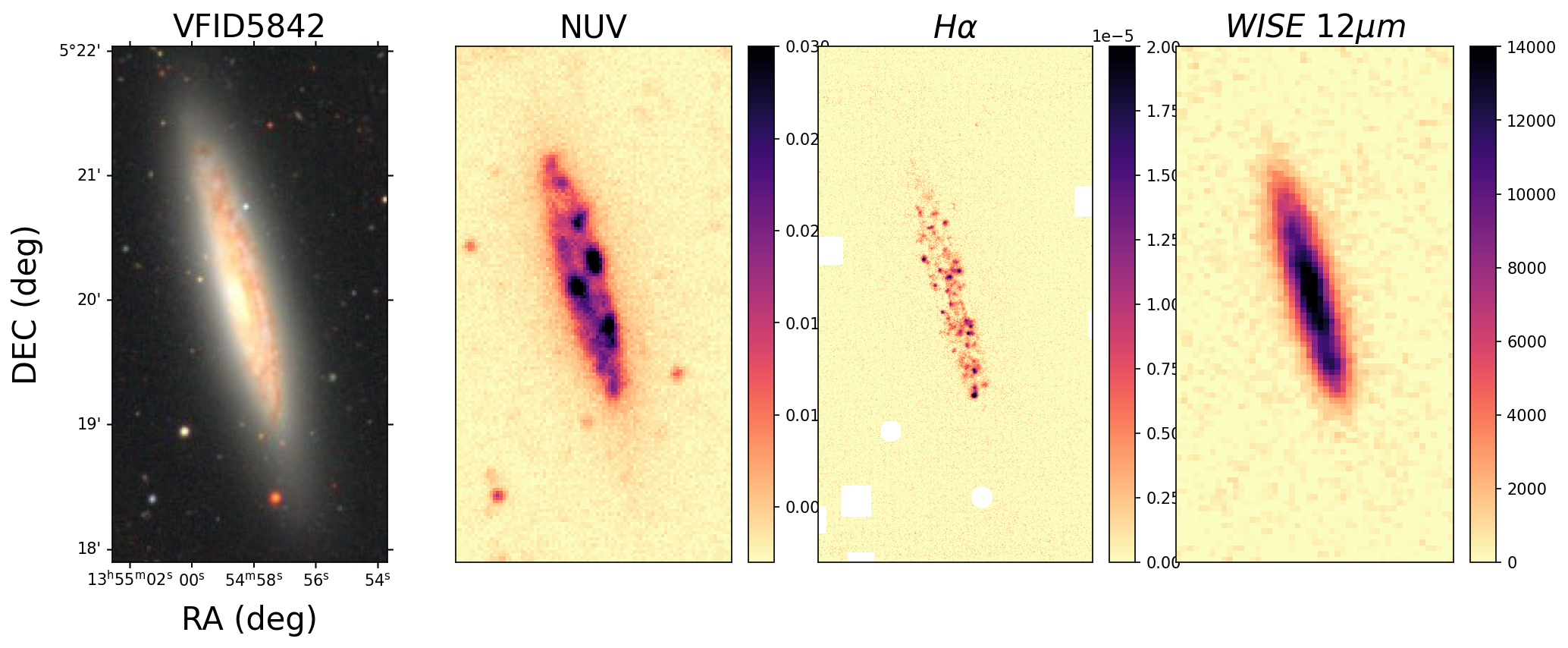}
        \includegraphics[width=\textwidth]{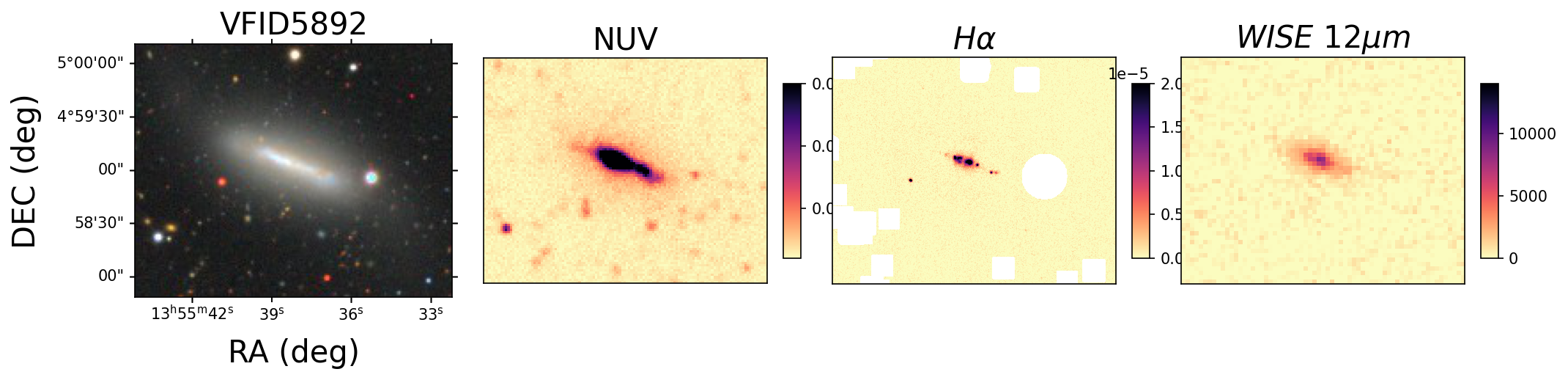}
    \caption{Multiwavelength SFR indicators NUV, \ha, and Wide-field Infrared Survey Explorer (WISE) 12\micron \ for (top) VFID5842 and (bottom) VFID5892.  Both galaxies show truncated \ha \ emission, and the NUV and IR show a similar extent to \ha. This indicates that dust is not causing the truncated \ha \ disks.}
    \label{fig:sfr_indicators}
\end{figure*}

To confirm that VFID5842 and VFID5892 have truncated star formation, we examine the Galaxy Evolution Explorer (GALEX) NUV and Wide-field Infrared Survey Explorer (WISE) W3 12~\micron \ images, which probe the unobscured and obscured star formation that occurs on slightly longer timescales.  We show the results in Figure \ref{fig:sfr_indicators}.  {While previous work shows that the spatial extents of UV and \ha \ can differ \citep[e.g.,][]{Boselli2006,Goddard2010}}, the NUV and 12~\micron \  images for VFID5842 and VFID5892 show a similar extent as \ha\, with a truncation radius that is smaller than the stellar disk.  Therefore, we find no evidence that the truncated \ha \ disks are due to dust obscuration.  Thus, it appears that the star formation in these galaxies is truly truncated relative to the stellar disk. 

\subsubsection{The Lowest-mass Galaxies: VFID5879, VFID5844, VFID5859}
\begin{figure*}
    \centering
        \includegraphics[width=\textwidth]{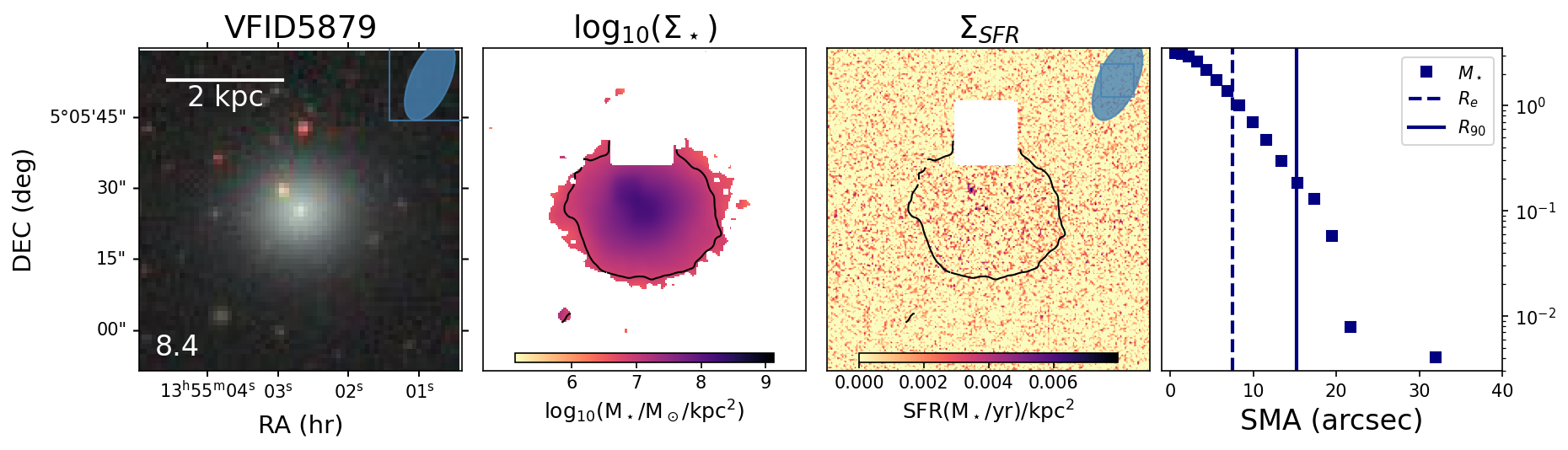}
        \includegraphics[width=\textwidth]{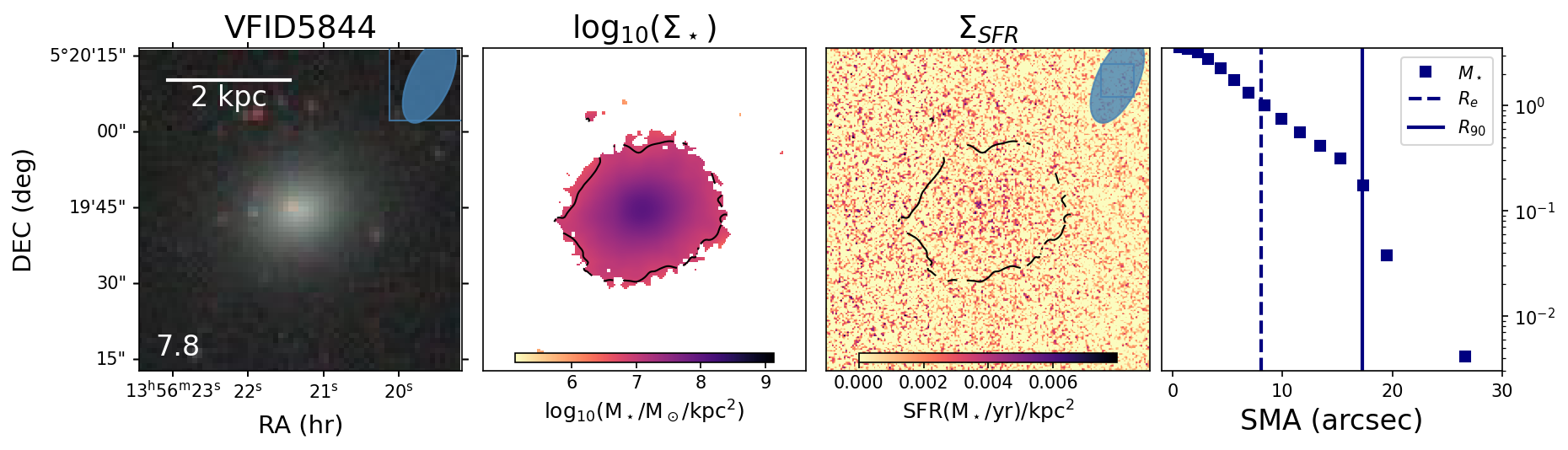}

         \includegraphics[width=\textwidth]{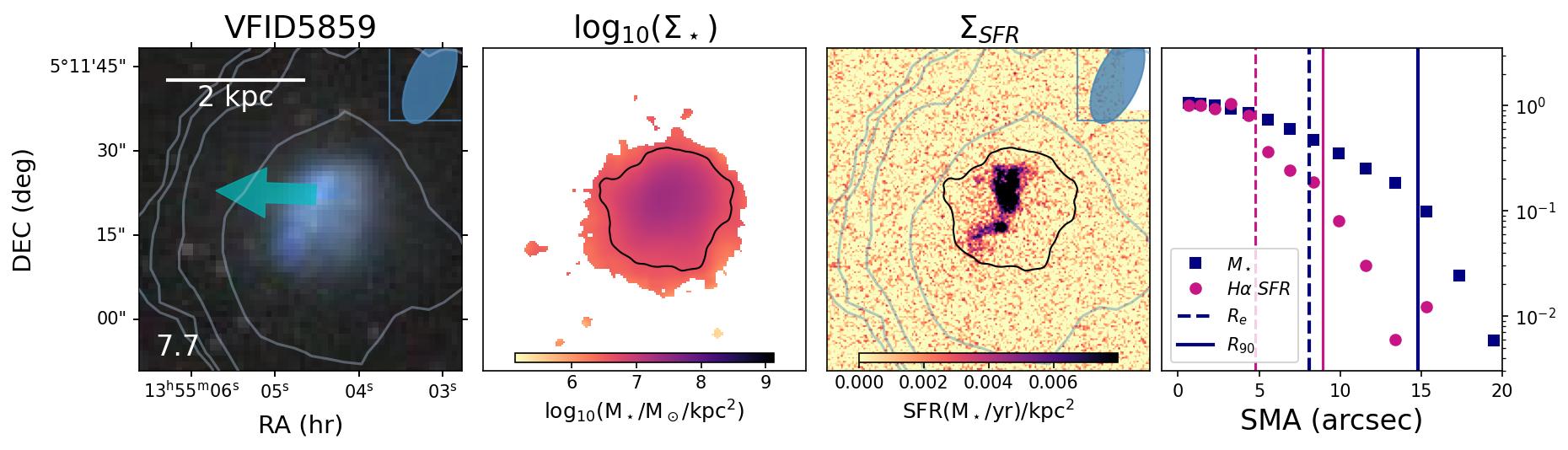}
    \caption{Results for $7.5 < \log_{10}(M_\star/M_\odot) < 8.5$ group galaxies, with galaxies ordered from top to bottom by decreasing stellar mass.  The left panel shows the Legacy $grz$ color image, with $\log(M_\star/M_\sun)$ written in the lower left and a scale bar in the top left.  The second panel shows the stellar mass map, with the color bar showing $\rm \log_{10}(M_\star/M_\sun/kpc^2)$.  The third panel shows  the \ha \ SFR map, with the color bar showing $\rm \log_{10}(M_\star/M_\sun ~yr^{-1})/kpc^2$.  The fourth panel shows the radial profiles of the stellar mass and SFR, normalized to unity in the central 5\arcsec. The black contours in the two middle panels show where  $\rm \log_{10}(M_\star/M_\sun/kpc^2) \approx 6.6$, and this is included to aid in comparing the spatial extent of the stellar mass and \ha \ SFR.  (Top and Middle) VFID5879 and VFID5844, two dwarf galaxies that show little to no \ha \ emission and no H~I.  The right panels for these two galaxies show the stellar mass profiles only because they do not have a significant amount of the \ha \ emission. (Bottom) VFID5859, a star-forming dwarf galaxy.  The MeerKAT H~I contours are shown in gray in the first and third panels, and the beam size is shown with the blue ellipse in the top right of the first and third panels.  The cyan arrow points to the group center.}
    \label{fig:mstar_dwarfs}
\end{figure*}

\begin{figure}[h]
    \centering
    \includegraphics[width=0.5\textwidth]{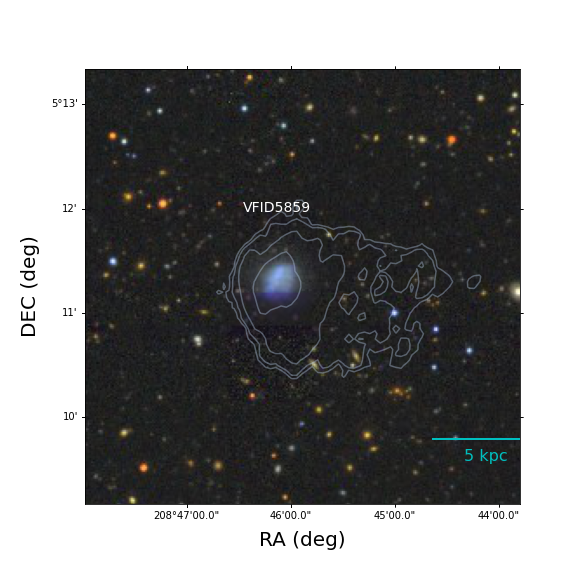}
    \caption{Legacy $grz$ color image of VFID5859 with MeerKAT H~I contours shown in gray.  This figure provides a zoomed-out image that includes the full H~I tail that is not captured in Fig. \ref{fig:mstar_dwarfs}.}
    \label{fig:VFID5859-zoom}
\end{figure}

In Figure \ref{fig:mstar_dwarfs}, we show the results for the three $7.5 < \log_{10}(M_\star/M_\odot) < 8.5$ galaxies that fall within the footprint of the \ha \ image. 
In the top and middle rows, we show the results for VFID5879 and VFID5844.  The \ha \ images reveal little to no star formation in these galaxies.  In the right panels, we show the radial distribution of the stellar mass but not the \ha \ SFR, as these galaxies have no significant \ha \ emission.  Neither VFID5879 nor VFID5844 was detected with MeerKAT. 
The bottom row of Figure \ref{fig:mstar_dwarfs} shows the results for VFID5859.  In contrast to VFID5879 and VFID5844, VFID5859 shows significant \ha \ emission.  Comparison of the \mstarim \ and \sfrim \ images shows that the star formation is asymmetric with respect to the stellar mass distribution and is concentrated on the east side of the galaxy.  We show the radial profiles of the stellar mass and SFR in the right panel, and the radii indicate that the SFR is more centrally concentrated than the stellar mass.  However, we note that $R_{50}$ and $R_{90}$ should be interpreted with caution in a case such as this, where the \ha \ emission is so asymmetric but the photometry is measured in elliptical apertures defined by the stellar mass distribution.  
The H~I emission shows extension to the west of the galaxy, in the opposite direction of where the \ha \ is concentrated.   The cyan arrow shows the direction to the group center, and the signatures in the star formation and H~I are consistent with the galaxy moving eastward through the intragroup medium.  The H~I is much more extended than the \ha \ emission, and it is difficult to capture the detail in both components in the same image.  We therefore show a zoomed-out image of VFID5859 in Figure \ref{fig:VFID5859-zoom} to show the full extent of the H~I tail.

\subsection{Global Properties of Group Galaxies}
\begin{figure*}
    \centering
    \includegraphics[width=0.48\textwidth]{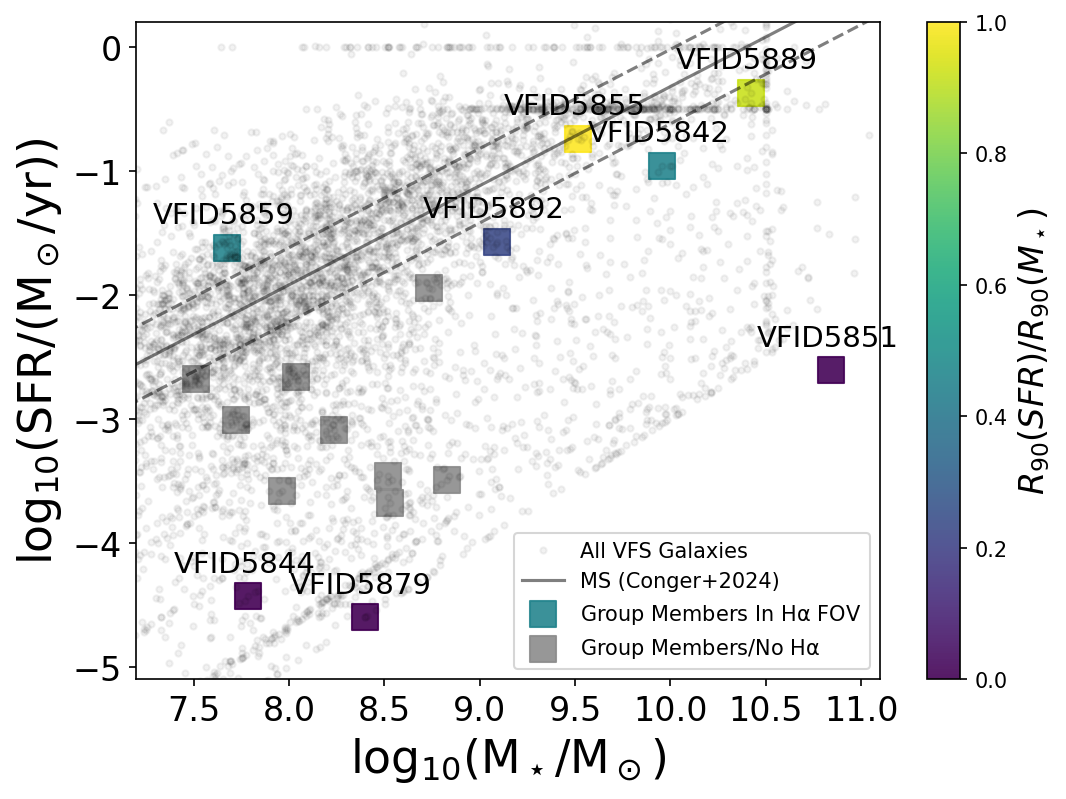}
        \includegraphics[width=0.48\textwidth]{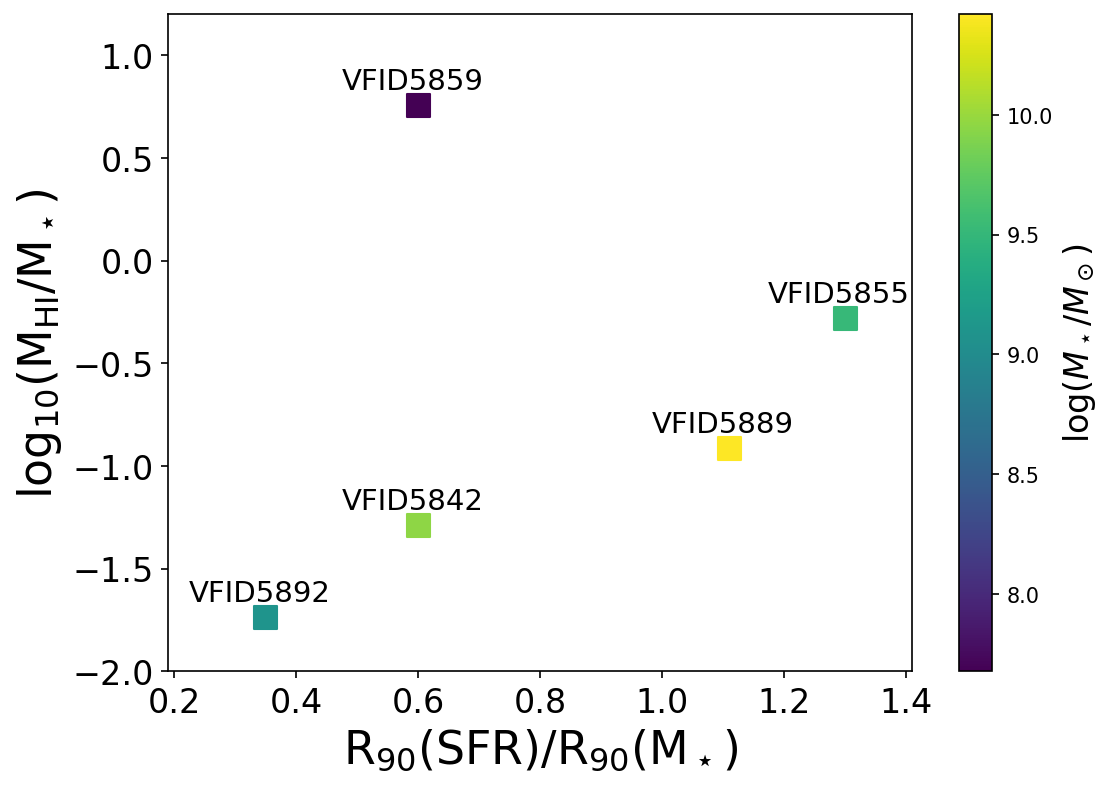}

    \caption{(Left) SFR vs. stellar mass for the full Virgo Filament Survey sample (gray) and the group members (squares).  For group members on the \ha \ image, the squares are color coded by the ratio of $R_{90}({\rm SFR})/R_{90}(M_\star)$.  The remaining group members are shown with the gray squares.  The solid line shows main-sequence fit from \citet{Conger2025}, and the dashed line shows the 0.3~dex scatter in the main sequence.
    (Right) Ratio of H~I mass to stellar mass vs. the size ratio of the \ha \ SFR profile to the stellar mass profile for the group galaxies with H~I detections from MeerKAT.}
    \label{fig:sfr-mstar}
\end{figure*}

In the left panel of Figure \ref{fig:sfr-mstar}, we show the SFR versus stellar mass for the full VFS sample (gray) and the NGC~5364 group members on the \ha \ image (squares). The squares are color coded by the
size ratio $R_{90}({\rm SFR})/R_{90}(M_\star)$ for the group members on the \ha \ image.  We set the ratio to zero for the two dwarf galaxies with no \ha \ emission (VFID5879 and VFID5844).  The solid line in Figure \ref{fig:sfr-mstar} shows the main-sequence fit from \citet{Conger2025}, and the dashed line shows the 0.3~dex scatter in the main sequence.  The two group galaxies with truncated \ha \ emission (VFID5892 and VFID5842) have SFRs that are below the main sequence.  VFID5859 lies above the main sequence, and this is a low-mass galaxy with asymmetric star formation and an H~I tail.  The galaxies with the lowest size ratios lie well below the main sequence (VFID5851, VFID5844, VFID5879). {While this may appear at face value to be expected, since smaller star-forming disks could result in lower SFRs, \citet{Conger2025} show that the ratio of WISE 12~\micron\ emission with the stellar mass does not correlate with position relative to the star-forming main sequence.  This could either be because the truncation of star formation in the outskirts of galaxies does not significantly affect the total, or it could be that the surface density of star formation is altered at the same time as the size in such a way as to preserve the total SFR. With our full \ha \ sample in a future paper we will determine the statistical relation, if any, between size ratio measured from \ha\ and the SFR.}

In the right panel of Figure \ref{fig:sfr-mstar}, we show the ratio of H~I mass to stellar mass versus the size ratio $R_{90}({\rm SFR})/R_{90}(M_\star)$.  The H~I masses are from MeerKAT (Ramatsoku et al, in preparation).  The color bar indicates stellar mass, which is not strongly correlated with either variable in the plot.
The two galaxies with the lowest values of $\log_{10}(M_{HI}/M_\star)$ (VFID5892, VFID5842) are the two galaxies with truncated \ha \ emission.  The two galaxies with H~I tails and unusual features in the \ha \ emission (VFID5855, VFID5859) have the highest values of $\log_{10}(M_{HI}/M_\star)$.    
A Spearman Rank Correlation test indicates that $\log_{10}(M_{HI}/M_\star)$ is not correlated with $R_{90}({\rm SFR})/R_{90}(M_\star)$ when considering all five galaxies ($r=0.56, p=0.32$).  However, as noted above, the  $R_{90}$ value for VFID5859 is not reliable given the asymmetry of the \ha \ emission.  If we run the Spearman Rank test without VFID5859, then the variables are highly correlated ($r=1.000, p=0.000$). We will be able to better test the strength of this correlation using the full VFS-\ha \ sample (Finn et al., in preparation), but similar trends have been found between H~I deficiency and the relative size of the star-forming and stellar disks \citep[e.g.,][]{Cortese2012, Fossati2013, Finn2018}.  

\citet{Castignani2022a} present integrated CO measurements of VFID5842, 5851, 5855, and 5889,
 and we list the $H_2$ deficiencies from \citet{Castignani2022a} in Table~\ref{tab:sample} for reference.  {There is no clear trend in $\rm H_2$ deficiency with $R_{90}({\rm SFR})/R_{90}({M_\star})$ or with the offset relative to the $SFR-M_\star$ main sequence.}  It is of interest that VFID5889 is the most $\rm H_2$ deficient of any of the galaxies, although it has only slightly lower star formation and is clearly a large star-forming spiral.  It may be that the $\rm H_2$ deficiency for VFID5889 is highly uncertain, as the aperture corrections needed to measure the total H$_2$ mass from the single-dish CO flux include many assumptions, some of which, e.g., the spatial profile of the molecular gas, may be incorrect.

\section{Discussion}
\label{sec:discuss}

The galaxies in the \ngc \ group exhibit a large range of \ha \ sizes and \ha \ morphologies.  Some have \ha \ emission that is similar in extent to the stellar mass, others have \ha \ emission that is significantly smaller than the stellar mass distribution, and some have no \ha \ emission at all.  In addition, the H~I morphology is truncated in some galaxies and exhibits clear signs of stripping in others.  As we discuss below in Section \ref{sec:processes}, these data suggest that multiple processes are working to disrupt the baryon cycle in these group galaxies.  Before discussing the group-related processes, we first discuss the potential influence of the Virgo Cluster on the evolution of the \ngc \ group galaxies.

\subsection{The Influence of the Virgo Cluster}

\begin{figure*}
    \includegraphics[width=0.5\textwidth]{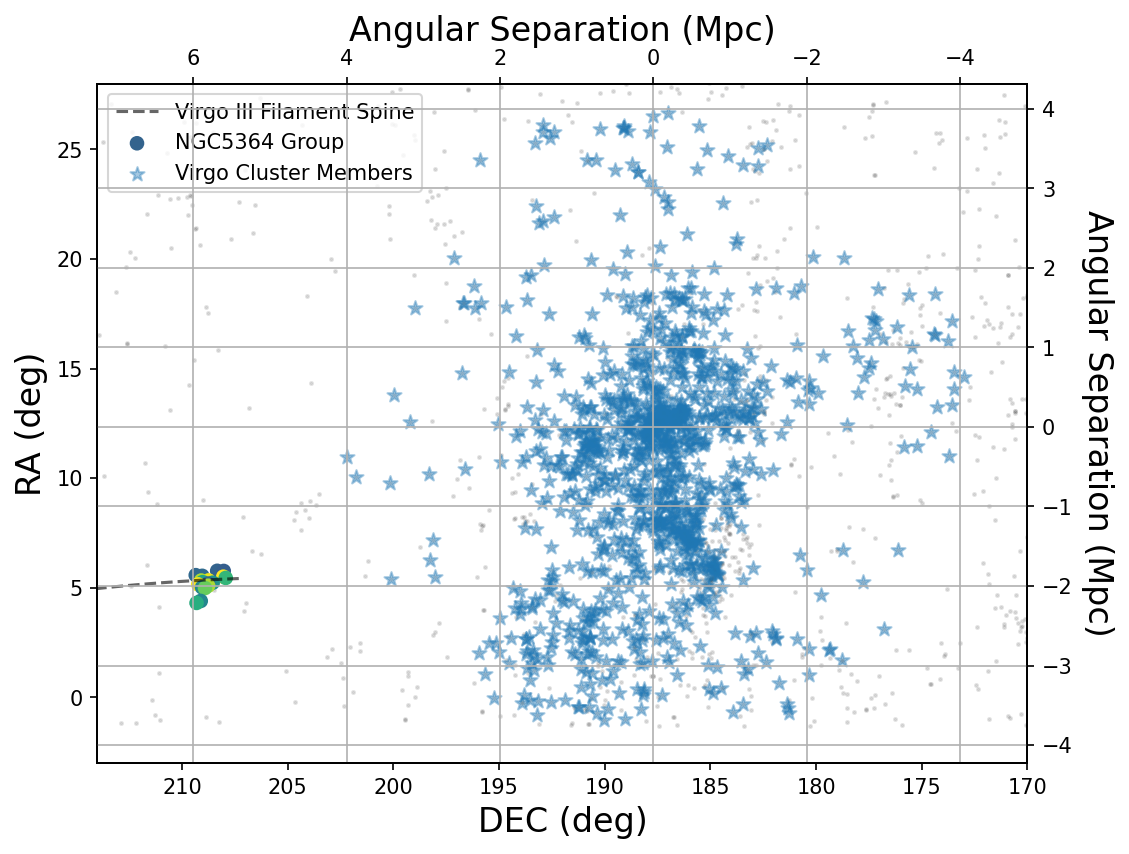}
        \includegraphics[width=0.5\textwidth]{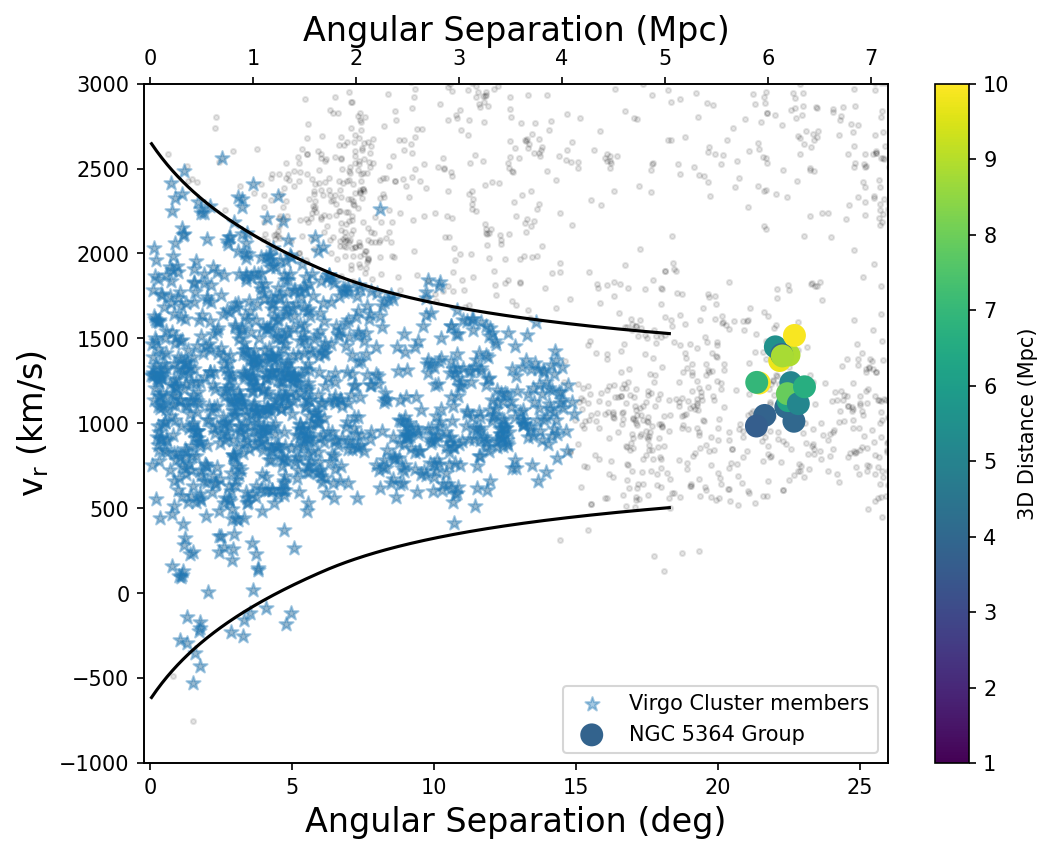}

    \caption{(Left) decl. vs. R.A. of NGC~5364 group members compared to Virgo Cluster members.  (Right) Recession velocity vs. projected angular separation from the center of Virgo.  Black lines show the model for the escape velocity from \citet{Castignani2022b}.}
    \label{fig:Virgoenv}
\end{figure*}

In Figure \ref{fig:Virgoenv}, we show the location of the NGC~5364 group with respect to the Virgo Cluster.  In the left panel, we show sky coordinates of the \ngc \ group with colored circles, and the color shows the 3D distance from Virgo.  Starred points show the Virgo cluster members from \citet{Castignani2022b}, and the dashed line shows the projected location of the Virgo~III filament spine.  The alternate $x$ and $y$-axis labels show how the angular distance translates into physical separation at the distance of the Virgo cluster, and from this we can infer that the group members are at least 6~Mpc away from the center of Virgo, which is significantly larger than Virgo's virial radius \citep[$R_{200}\approx 1$~Mpc;][]{McLaughlin1999}.  We note that the virial radius may not be the best to describe the sphere of influence of Virgo, as the distribution of galaxies is clearly not spherically symmetric but is instead elongated in the north-south direction.

To assess the possibility of hydrodynamic effects from the Virgo Cluster, we attempt to estimate the density of the intracluster medium (ICM) at a projected radius of $6R_{200}$ from the Virgo cluster using the eROSITA-derived ICM density profile from \citet{McCall2024}.  They only measure the profile out to $3R_{200}$ and give a power-law fit to the density profile that is only valid out to $0.4R_{200}$.  Toward the east and south-east the profiles flatten out beyond $R_{200}$.  This flattening is attributed to the ``eROSITA Bubble" \citep{Predehl2020}, which extends far out of the plane of the Milky Way and is attributed to energy injected into the galactic halo from the galactic center.  Nonetheless, as a rough estimate of the ICM density, we extrapolate the density profile out to $6R_{200}$, with the understanding that this is very uncertain.  The projected ICM density at this radius is $\sim$10 times lower than at $R_{200}$.  This is roughly 1.5~dex less than the intragroup medium density expected for groups of this mass (Section~\ref{sec:RPS_calculation}).  We therefore feel that any ram pressure effects on galaxies in the \ngc\ group are dominated by the intragroup medium and not by interactions with Virgo's ICM.  This supposition is supported by the observation that the passive fraction of galaxies drops monotonically toward the outskirts of nearby clusters and is nearly at the field value by $3\times R_{200}$.  No studies measure the passive fraction out to $6\times R_{200}$ in $z\sim 0$ clusters, but \citet{Pintos-Castro2019} find that the passive fraction around clusters at $0.3<z<0.5$ reaches the field value by $\sim$2$\times R_{200}$.  This implies that clusters themselves do not quench galaxies significantly beyond $\sim$2$\times R_{200}$.

A similar result is illustrated in the right panel of Figure \ref{fig:Virgoenv}, which shows a phase-space diagram with recession velocity versus projected angular separation (in both degrees and Mpc).  The caustics are from \citet{Castignani2022b} and show the radial dependence of the escape velocity as computed using the prescription from \citet{Jaffe2015}.  The group has a low velocity offset with respect to the  Virgo cluster.  In addition,  the \ngc \ galaxies are not in the regions where one would expect to see backsplash galaxies -- galaxies that have already passed through the center of the cluster -- as the backsplash population in clusters dominates between 1 and 2 virial radii \citep{Balogh2000}. Therefore, we conclude that the observed trends in star formation and H~I properties are not due to the Virgo cluster.

\subsection{Observing the Disruption of the Baryon Cycle in Galaxy Groups}
\label{sec:processes}
{There are multiple ways that the baryon cycle in group galaxies can be disrupted \citep{George2018,Poggianti2019, Boselli2022}.  Thanks to the high densities and low relative velocities, tidal interactions are common in groups and can remove gas from the halos and outskirts of galaxies.  For example, \citet{Williamson2016} used hydrodynamic simulations to show that tidal interactions within a group can preferentially remove gas in the outer parts of disks in dwarf galaxies, which can then lead to truncated disks.  These tidal interactions can also result in extraplanar star-forming gas \citep{Stein2017}.   
While ram pressure stripping in clusters can remove dense molecular gas from galaxies \citep{Vollmer2008}, purely hydrodynamical effects can also occur via a galaxy's passage through the intragroup medium \citep{Roberts2021}.  In its weak manifestation, hydrodynamic stripping from the intragroup medium can remove gas from the hot halo and also strip the loosely bound gas on the edges of galaxy disks.  
The intragroup medium can also interfere with gas recycling within a galaxy.  For example, galactic fountains that are ejected from galaxies can extend to large distances (60-100~kpc) and up to 30\% of that material can be reaccreted onto galaxies \citep{Oppenheimber2008}.  \citet{Bahe2015} found that this gas can also be stripped if it travels far enough away from the disk, thus removing a major source of gas replenishment and causing a reduction in the SFR \citep{Oman2021}.}  These physical mechanisms can also interplay in complex ways.  For example, tidal interactions and supernova feedback can make the gas more susceptible to ram pressure stripping \citep{Marasco2016, Serra2023}.  Given how conducive the group environment is to multiple processes, trying to identify a single dominant mechanism is not well motivated. 

Indeed, the main spiral (VFID5889) and the spiral located to its southwest (VFID5892) show evidence of both gravitational and hydrodynamical interactions. VFID5889 has a significantly lopsided stellar, H~I, and \ha\ disk, which can be caused by weak tidal interactions \citep{Walker1996,Zaritsky1997,Rudnick1998}.  It also has very extended diffuse stellar emission, but the H~I only extends to the edge of the spiral arms.  Isolated galaxies often have H~I disks that extend well beyond $r_{25}$ of the stellar light \citep[e.g.,][]{Bigiel2008}, and the lack of extended H~I emission in VFID5889 could indicate that the outer H~I has been removed.   VFID5892 shows signs of a tidal interaction with VFID5889, as evidenced by a warped stellar disk.  The gas and SFR distribution in VFID5892 have also been modified.  The H~I disk is truncated, and this galaxy has the highest H~I deficiency and the smallest size ratio of \ha\ to stars of the entire sample (Figure~\ref{fig:sfr-mstar}).  The \ha\ and H~I disks are also off-center with respect to the stellar disk (Figure~\ref{fig:mstar_spirals}), which is what is naively expected from hydrodynamic processes.  It is not clear if such a feature can also be caused by tidal interactions.

Two galaxies, VFID5855 and VFID5859, show asymmetric H~I distributions that are compressed on one side and extended on the other (Figs.~\ref{fig:mstar_spirals} and \ref{fig:mstar_dwarfs}), forming tails that match the qualitative expectation for ram pressure stripping.  Indeed, the H~I contours for VFID5855 show streamers of H~I trailing behind the disk.  VFID5855 also shows extraplanar \ha\ emission, and the H~I contours on that side of the disk trace the \ha\ and are curved in the direction of the tail on the leading edge of the disk.  The tails for VFID5855 and VFID5859 point in opposite directions.  It may also be possible that the two galaxies recently experienced an interaction and that this partly explains the extraplanar \ha \ emission at the southern end of VFID5855.  The distortion in the stars of VFID5855 is not strong enough to conclusively establish the presence of an interaction using published diagnostic criteria \citep{Vollmer2005,Vulcani2021}.  
We will discuss the stripping in VFID5855 in detail in Section \ref{sec:RPS_calculation}, but if an interaction is indeed occurring, it could 
boost the susceptibility of the gas to ram pressure stripping, as discussed in \citet{Serra2023}.  

The only galaxy with a clearly truncated H~I profile and \ha\ disk, but with no evidence of tidal interaction, is VFID5842.  This galaxy has a symmetric distribution of stellar mass, H~I, and SFR, but the \ha \ and H~I disks are clearly truncated.  As noted in Section \ref{sec:results_spatdist}, while $R_{50}({\rm SFR})/R_{50}(M_\star)\sim~1$, the $R_{90}$ values for SFR and the stellar mass differ significantly.  Within the truncation radius, the stellar and SFR profiles have a similar slope, indicating naively that whatever affected the gas distribution preferentially modified the outskirts of the galaxy.  We will discuss this object and its susceptibility to RPS in detail in Section \ref{sec:RPS_calculation}.

The elliptical VFID5851 has little H~I and \ha\ emission, and the \ha\ profile is significantly steeper than that of the stars.   
The origin of the observed gas is unclear. 
Work on the properties of nearby early-type galaxies has revealed that the H~I detection fraction depends strongly on environment \citep{Serra2012} but that the CO detection rate is independent of environment \citep{Young2011}.  The presence of \ha\ emission in this galaxy could have many sources, including mass loss from existing stellar populations or accretion from the cosmic web, though it is not clear how the latter is modified by the low-mass group environment \citep{Rudnick2017}.

The remaining two galaxies within the \ha \ FOV are the dwarf galaxies VFID5879 and VFID5844, and these are not detected in \ha\ or H~I (Fig. \ref{fig:mstar_dwarfs}).  Both galaxies are detected in the GALEX NUV images.  If this NUV emission is due to the most recent episode of massive star formation, then this indicates fairly recent  quenching \citep[longer than \ha \ timescale of 10~Myr but less than NUV timescale of 100~Myr, e.g.,][]{Kennicutt1998}. There are no signs of tidal interaction in these galaxies.  They are low mass ($\log_{10}(M_\star/M_\sun) < 8.5$), and therefore ram pressure stripping could be effective in quenching these galaxies, as could starvation.   

The passive fraction for the eight galaxies in the \ha\ image of this group is $3/8 = 0.375$.  This is significantly higher than the passive fraction for isolated galaxies, which is at most 0.05 for the highest masses and falls to 0.01 at $\log(M_\star/M_\odot)<9.5$ \citep{Geha2012}.  Given the small numbers in our sample, we can use binomial statistics to rule out the field value at the 90$-$99\% level, depending on the mass limit adopted.  When comparing to the passive fractions of satellite galaxies, the passive fraction of the $\log(M_{halo}/M_\odot)=12.7$ \ngc \ group falls between those of Milky Way analogs with $\log(M_{halo}/M_\odot)=12-12.3$, which are almost entirely star-forming \citep{Geha2017}, and groups with $\log(M_{halo}/M_\odot)=13-14$, which have passive fractions at these masses of $\sim 0.5$ \citep{Baxter2021}.  The observed passive fraction of 0.375 implies that the \ngc \ group may indeed be conducive to galaxy quenching.

\subsection{Quantifying the Role of Ram Pressure Stripping}
\label{sec:RPS_calculation}

In this section, we use the classical description of ram pressure stripping \citep{Gunn1972} to explore its potential role in the ongoing evolution of the four disk-dominated galaxies (VFID5842, VFID5855, VFID5851, VFID5892).
Several other studies have found ram pressure candidates in groups based on the presence of X-ray tails \citep[e.g.,][]{Sivakoff2004, Rasmussen2006}, H~I tails \citep[][]{Roberts2021,Maina2022}, and optical signatures of stripping, such as jellyfish galaxies \citep[e.g.,][]{Vulcani2018, Vulcani2021, Kolcu2022}. \citet{Vulcani2018} also find multiple signatures of environmental processing, including truncated star-forming disks, in four galaxies
that are part of the same group ($z = 0.06359$). However, their group has a mass of $\sim5 \times 10^{13}~M_\odot$, which is at least a factor of 5 greater than the mass of the \ngc \ group.
Therefore, we want to explore the potential influence of RPS in a group with much lower halo mass.

The strength of ram pressure stripping depends on both the density of the intracluster or intragroup medium (hereafter ICM for simplicity and to avoid confusion with IGM$\equiv$ intergalactic medium)  and the velocity of the galaxy with respect to this hot intergalactic gas: 
\begin{equation}
    P_{RPS} = \rho_{ICM} \Delta v^2,
\end{equation}
where $P_{\rm RPS}$ is the pressure exerted by the ICM, $\rho_{\rm ICM}$ is the density of the ICM, and $\Delta v$ is the velocity of a galaxy with respect to the ICM.
The effectiveness of RPS also depends on the mass and structure of the infalling galaxy because this determines the galaxy's ability to keep its gas gravitationally bound.
According to the classical treatment of ram pressure stripping presented in \citet{Gunn1972}, the restoring force per area in a disk galaxy is given by:
\begin{equation}
    F/A = 2 \pi G\Sigma_* \Sigma_g
    \label{eqn:restoring_force}
\end{equation}
where $\Sigma_*$ is the surface density of stars and $\Sigma_g$ is the surface density of gas.  
When ram pressure exceeds this restoring force, the gas can be pushed off the disk.  
We note that the effectiveness of RPS also depends on the details of a galaxy's orbit \citep[e.g.,][]{Tonnesen2009, Tonnesen2019} as well as the angle of the galaxy disk with respect to the wind \citep[e.g.,][]{Vollmer2021}.  However, we do not consider these additional factors here for the sake of simplicity. 

If a galaxy shows a clear truncation of either the \ha \ or H~I emission, then we can use the outer radius to estimate the density of the ICM if the truncation is in fact due to RPS.  
Specifically, if we assume that the outer disk gas has been removed by ram pressure, then the truncation radius, $R_{trunc}$, indicates the location where the restoring force per unit area is equal to the ram pressure:
\begin{equation}
\label{eqn:rps}
\rm 
    \rho_{IGM} \Delta v^2 =  2 \pi G\Sigma_*(R_{trunc}) \Sigma_g(R_{trunc})
\end{equation}
At smaller radii, the restoring force exceeds ram pressure, and at larger radii, ram pressure dominates.  We calculate the stellar and gas mass surface densities by assuming exponential profiles for the stellar and gas distributions \citep[e.g.,][]{Domainko2006, Jaffe2015}:
\begin{equation}
\rm 
  \Sigma_*(R_{trunc})  = \frac{M_\star}{2 \pi R_{d,stars}^2} e^{-R_{trunc}/R_{d,stars}},
\end{equation}
and
\begin{equation}
\rm 
  \Sigma_g  = \frac{M_{HI} + M_{H_2}}{2 \pi R_{d,gas}^2}e^{-R_{trunc}/R_{d,gas}},
\end{equation}
where $R_{d,stars}$ is the scale length of the stellar disk and  $R_{d,gas}$ is the scale length of the gas \citep[see also ][]{Gullieuszik2020}.  We have estimated the half-light radius $R_{50}$ in Section \ref{sec:results_spatdist}, and the half-light radius and scale length are related by $R_{50} = 1.67 R_{d}$.
We assume the scale length of the gas disk is 1.7 times greater than the scale length of the stellar disk \citep[$ R_{d,gas} = 1.7 R_{d,stars}$; e.g.,][]{Cayatte1994, Jaffe2015}.
It is in theory possible to use the \ha \ emission to estimate the extent of the molecular gas using the Schmidt law \citep{Bigiel2008}.  However, H~I can extend well beyond the \ha \ emission even for galaxies undergoing RPS, as for VFID5855 (Fig. \ref{fig:mstar_spirals}), and \ha \ would underestimate the gas scale length in this case. This could even be true for truncated galaxies, depending on exactly how the gas is being modified. In Castignani et al. (2025, in preparation), we will examine the detailed spatial profiles of the molecular gas, atomic gas, and star formation for a set of galaxies, including some of those in this paper.

To estimate the restoring force,
we first determine the outer radius of each galaxy, which we hereafter call the truncation radius.
For VFID5889, VFID5842, and VFID5892, the \ha \ and H~I are similar in spatial extent, and we use the outer radius of this emission as the truncation radius.  For VFID5855, the H~I extends beyond the \ha \ emission, and so we use the extent of H~I along the major axis to determine the truncation radius.  
To estimate the surface density of atomic and molecular gas, we use the H~I masses from MeerKAT observations (Ramatsoku et al., in prep) and 
the molecular gas mass estimates from \citet{Castignani2022a}.  
We then combine the truncation radius, the stellar mass, and the gas mass according to Equation \ref{eqn:restoring_force} to estimate the restoring force per area.  

\begin{figure}
    \centering
    \includegraphics[width=0.5\textwidth]{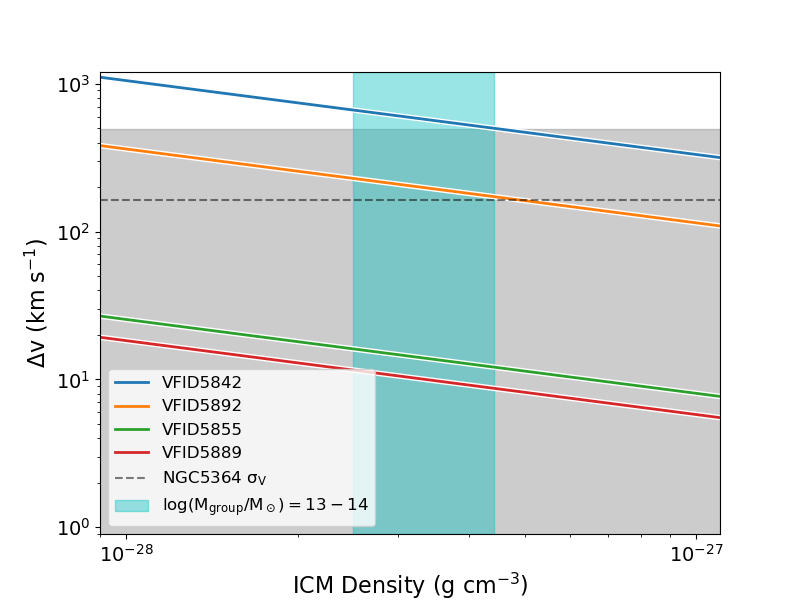}
    \caption{To illustrate the potential influence of ram pressure stripping, we plot the relative velocity of galaxy with respect to group center vs. the density of the ICM.  The solid diagonal lines show lines of constant pressure, and each line reflects the pressure required to remove mass at the observed truncation radius.  The horizontal dashed line shows the group velocity dispersion from \citet{Kourkchi2017}, and the gray shaded region shows $+3\sigma_V$.  The shaded cyan region shows the approximate ICM density measured for a sample of $13 < \log_{10}(M/M_\odot) < 14$ galaxy groups \citep{Sun2009} 
    as presented in \citet[][see their Fig. 1]{Boselli2022}.  
    }
    \label{fig:RPS-calculation}
\end{figure}

We show the results in Figure \ref{fig:RPS-calculation}, where we plot the restoring force for each galaxy relative to the two factors that affect the strength of ram pressure stripping: $\Delta v$ and $\rho_{IGM}$.  
For a particular galaxy, any combination of $\Delta v$ and $\rho_{IGM}$ that falls along its line indicates conditions where ram pressure is equal to the restoring force, or where RPS can produce the observed truncation radius.  The regions in $\Delta v$$-$$\rho_{IGM}$ space that fall below a galaxy's line indicate conditions where RPS is insufficient to produce the observed truncation radius.
We can constrain the likely range of $\Delta v$ using the group's velocity dispersion ($\approx$150~$\rm km~s^{-1}$, see Sec. \ref{sec:ngcgroup}), which we show with the horizontal dashed line.  The horizontal shaded region shows $\Delta v~\le~3\sigma$ to denote the region along the $\Delta v$ axis where we expect most group members to lie.  

We use information compiled by \citet{Boselli2022} to estimate the range of ICM densities for galaxy groups.  The ICM density decreases as the distance from the group/cluster center increases.  We calculate the projected distance of the four galaxies from the group center and find an average distance of approximately $0.4 R_{\rm virial}$, where $R_{\rm virial} = 222$~kpc comes from \citet{Kourkchi2017}.  From \citet{Boselli2022} Figure 1, the 
density of the ICM for a sample of 43 galaxy groups \citep{Sun2009} ranges from $\rho_{ICM} = 2.5 - 4.4 \times 10^{-28}~\rm~g~cm^{-3}$. 
We note that the groups in that study have $13 < \log_{10}(M/M_\odot) < 14 $, so they are likely more massive than the NGC~5364 group.  
We show the range of ICM densities expected at $0.4 R_{\rm virial}$ for these galaxy groups with the cyan region in Figure \ref{fig:RPS-calculation}.  
For comparison, the central gas mass density for Coma 
would be off the right edge of the plot at $\rm \rho_{ICM} =  7.8 \times 10^{-27}~g~cm^{-3}$ \citep{Gunn1972}.

The calculations of restoring force shown in Figure~\ref{fig:RPS-calculation} are based on a number of assumptions and should only be considered as approximate.  However, the comparison of restoring force and ram pressure is instructive.   
If the curves in the figure reside where the gray and cyan regions overlap, then RPS could be effective.  
The pressures required for VFID5892, VFID5855, and VFID5889 are consistent with the range of $\Delta v$ and $\rho_{ICM}$ values expected for this group. The blue curve in Figure \ref{fig:RPS-calculation} shows the restoring force at the truncation radius for VFID5842, and this galaxy requires both a high relative velocity and high ICM density to explain the truncation radius using RPS.  
Thus, the truncation observed in VFID5842 implies an external pressure that is at the upper edge of what is expected from the classical treatment of RPS.  However, it should be noted that the effectiveness of RPS can be increased by density variations in the ISM \citep[e.g.,][]{Quilis2000, Tonnesen2009} and ICM \citep[e.g.,][]{Tonnesen2008}. In addition, RPS can be enhanced due to the group's location within a filament.  For example, \citet{Bahe2013} show that in simulations, RPS can be enhanced by a factor of 100 in filaments compared to isolated galaxies.  
On the other hand, \citet{Zakharova2024} attributed gas depletion processes of galaxies in VFS filaments primarily to the groups embedded in the filaments; they did not find evidence of filaments alone affecting galaxy gas contents.  Without a larger sample of galaxies in groups and filaments, it is not possible to more precisely isolate the additional effect that filaments have on group galaxies.  Likewise, there are not enough direct studies of the intragroup medium for groups inside and outside of filaments to determine if filaments can enhance the RPS in groups.  With this in mind, the main takeaway from Figure \ref{fig:RPS-calculation} is that RPS from the intragroup medium is a reasonable explanation for the truncated H~I and \ha \ profiles observed in most of the disk-dominated group galaxies, perhaps with the exception of VFID5842.

\subsection{Additional Processes That Can Truncate Disks}
\label{sec:theory}

While we explore in detail the potential role of RPS in creating the truncated \ha \ and H~I disks in Sec. \ref{sec:RPS_calculation}, from a theoretical perspective, several additional physical processes can lead to truncated \ha\ and H~I disks in the group environment. 
 For example, as mentioned in Section \ref{sec:processes}, simulations show that tidal interactions within a group can lead to truncated disks \citep{Williamson2016}.
 Additionally, it is possible that starvation can reduce the size of the star-forming disk.  In the most straightforward implementation of starvation, accretion is halted and the remaining gas is consumed through star formation. The gas content should reduce uniformly across the disk as a consequence of the relatively constant depletion time of 2$-$3~Gyr throughout galactic disks \citep{Bigiel2008}.  This could in principle result in a truncation of the star-forming disk as the gas near the galaxy outskirts drops below the threshold for star formation of 3-4 $M_\odot~{\rm pc}^{-2}$ \citep{Leroy2008}.  However, this simple scenario ignores the impact of recycling and redistribution of gas in the disk.  For example, the GAEA simulations \citep{Xie2017,Xie2020,DeLucia2024} divide each galaxy into a set of annuli and track the consumption, recycling, expulsion, and stripping, as well as the redistribution of gas among these annuli.  The gas that is ejected via a galactic fountain is retained, and because the ejected gas has low angular momentum, it is added back at small galactocentric radii.  When galaxies in GAEA become satellites, they are cut off from accretion of new gas; the new gas should have higher angular momentum than gas accreted at earlier cosmic epochs and is therefore accreted at preferentially larger radii.  The net result of both outflows and accretion makes the gas, and hence star formation, more concentrated in the presence of starvation.  While this scenario has not been thoroughly explored through zoom-in hydrodynamic simulations except for the work by \citet{Kawata2008}, the implication is that it is possible to have a reduction in the \ha \ and H~I profiles just in the presence of starvation.

\section{Summary}

\label{sec:summary}
We present the first results from the VFS-\ha, an \ha \ imaging survey of over 600 galaxies that reside in the large-scale structure around the Virgo Cluster. 
In this paper, we focus on the \ngc \ group,  a low-mass group located at the western end of the Virgo~III filament and over 6~Mpc from the center of the Virgo Cluster.  
We analyze the \ha \ star-formation maps for eight group galaxies that were obtained with the WFC at the 2.5~m Isaac Newton Telescope, and we combine these with new high-resolution MeerKAT imaging of the H~I in these group galaxies.
The combination of \ha\ and high-resolution, sensitive H~I imaging is an extremely powerful tool for understanding the way different gas phases are affected in the intermediate-density environments that host most galaxies.  

We find that eight group members within the \ha \ image footprint show a range of properties.  We summarize these results in order of decreasing stellar mass.
\begin{itemize}
  \item The most massive galaxy in the group (VFID5851, an elliptical) shows diffuse, smooth, extended \ha \ and H~I in the central region (Fig. \ref{fig:mstar_massive}).
    \item The most massive spiral (VFID5889) shows \ha \ over the entire stellar disk.  The H~I emission is comparable in extent to the \ha \ emission but does not extend beyond the stellar disk (Fig. \ref{fig:mstar_massive}). Both the \ha\ and H~I are lopsided, as is the stellar disk.
    
    \item Two galaxies in this group show some evidence of mutual tidal interaction (VFID5889 and VFID5892), as indicated by a lopsided stellar distribution in VFID5889 and a stellar warp in VFID5892.  

    \item Two additional galaxies have clear H~I tails indicative of ram pressure stripping and have potentially just completed a close interaction.  One of these galaxies 
    (VFID5855; $ \log(M_\star/M_\sun)=9.5$) has extraplanar H~I and \ha \ (Fig. \ref{fig:mstar_spirals}).  Both of these galaxies appear to have H~I that is compressed on the side opposite the tail.   
    
    \item There are two galaxies (VFID5842, VFID5892) with clear H~I and \ha\ truncation.  One of these galaxies is likely interacting with a nearby large spiral galaxy, but the galaxy with the most significant truncation has no evidence of tidal interaction (Fig. \ref{fig:mstar_spirals}).
  
    \item Two low-mass galaxies (VFID5879, VFID5844) show no detectable \ha \ and H~I emission (Fig. \ref{fig:mstar_dwarfs}).
    
\end{itemize}

Despite the group's low mass ($\log(M/M_\odot) \sim 13$), we find multiple signatures of environmental processing that are disrupting the baryon cycle in a variety of ways.
In addition, we find that the ratio of H~I mass to stellar mass is correlated with the normalized size of the \ha \ disk.  We examine the quantitative relevance of RPS and find that it is sufficient to explain the truncated \ha \ and H~I and emission observed in all but one of the disk-dominated galaxies.  
The \ngc \ group in this paper is significantly lower in mass than those with prior RPS detections and/or truncated star-forming disks.  Thus, our results imply the ubiquitous modification of multiple phases of the gas content in group galaxies. 

In future work, we will analyze our entire MeerKAT H~I sample (Ramatsoku et al. 2025, in preparation) and the full \ha\ imaging survey (Finn et al. 2025, in preparation)  The latter contains comparable \ha\ images for $>600$ galaxies in the filaments and groups surrounding Virgo, $\sim$240 of which have integrated CO fluxes and H~I fluxes.  
Analysis of \ha \ images for an additional $\sim$40 groups ($12< \log_{10}(M/M_\odot) < 13.7$) in the VFS will help quantify the prevalence of these signatures in group galaxies.

\begin{acknowledgements}

R.A.F. gratefully acknowledges support from NSF grants AST-1716657 and AST-2308127 and from a NASA ADAP grant 80NSSC21K0640.

G.H.R. acknowledges the support of NASA ADAP grant 80NSSC21K0641, and NSF AAG grants AST-1716690 and AST-2308127. G.H.R. also acknowledges the hospitality of Hamburg Observatory, who hosted him during parts of this work. 

The authors thank the International Space Sciences Institute (ISSI) in Bern, Switzerland who hosted collaboration meetings as part of the ISSI COSWEB team. They also thank the Institute for Fundamental Physics of the Universe (IFPU) in Trieste, Italy for hosting a group workshop.  G.H.R., R.A.F., and B.V. thank Padova Observatory for hosting them during a team meeting.  

D.Z. and B.V. acknowledge support from the INAF Mini Grant 2022 “Tracing filaments through cosmic time” (PI Vulcani).

G.C. acknowledges the support from the Next Generation EU funds within the National Recovery and Resilience Plan (PNRR), Mission 4 - Education and Research, Component 2 - From Research to Business (M4C2), Investment Line 3.1 - Strengthening and creation of Research Infrastructures, Project IR0000012 – “CTA+ - Cherenkov Telescope Array Plus”.
 
The Legacy Surveys consist of three individual and complementary projects: the Dark Energy Camera Legacy Survey (DECaLS; Proposal ID \#2014B-0404; PIs: David Schlegel and Arjun Dey), the Beijing-Arizona Sky Survey (BASS; NOAO Prop. ID \#2015A-0801; PIs: Zhou Xu and Xiaohui Fan), and the Mayall z-band Legacy Survey (MzLS; Prop. ID \#2016A-0453; PI: Arjun Dey). DECaLS, BASS and MzLS together include data obtained, respectively, at the Blanco telescope, Cerro Tololo Inter-American Observatory, NSF’s NOIRLab; the Bok telescope, Steward Observatory, University of Arizona; and the Mayall telescope, Kitt Peak National Observatory, NOIRLab. Pipeline processing and analyses of the data were supported by NOIRLab and the Lawrence Berkeley National Laboratory (LBNL). The Legacy Surveys project is honored to be permitted to conduct astronomical research on Iolkam Du’ag (Kitt Peak), a mountain with particular significance to the Tohono O’odham Nation.

NOIRLab is operated by the Association of Universities for Research in Astronomy (AURA) under a cooperative agreement with the National Science Foundation. LBNL is managed by the Regents of the University of California under contract to the U.S. Department of Energy.

This project used data obtained with the Dark Energy Camera (DECam),  
which was constructed by the Dark Energy Survey (DES) collaboration. Funding for the DES Projects has been provided by the U.S. Department of Energy, the U.S. National Science Foundation, the Ministry of Science and Education of Spain, the Science and Technology Facilities Council of the United Kingdom, the Higher Education Funding Council for England, the National Center for Supercomputing Applications at the University of Illinois at Urbana-Champaign, the Kavli Institute of Cosmological Physics at the University of Chicago, Center for Cosmology and Astro-Particle Physics at the Ohio State University, the Mitchell Institute for Fundamental Physics and Astronomy at Texas A\&M University, Financiadora de Estudos e Projetos, Fundacao Carlos Chagas Filho de Amparo, Financiadora de Estudos e Projetos, Fundacao Carlos Chagas Filho de Amparo a Pesquisa do Estado do Rio de Janeiro, Conselho Nacional de Desenvolvimento Cientifico e Tecnologico and the Ministerio da Ciencia, Tecnologia e Inovacao, the Deutsche Forschungsgemeinschaft and the Collaborating Institutions in the Dark Energy Survey. The Collaborating Institutions are Argonne National Laboratory, the University of California at Santa Cruz, the University of Cambridge, Centro de Investigaciones Energeticas, Medioambientales y Tecnologicas-Madrid, the University of Chicago, University College London, the DES-Brazil Consortium, the University of Edinburgh, the Eidgenossische Technische Hochschule (ETH) Zurich, Fermi National Accelerator Laboratory, the University of Illinois at Urbana-Champaign, the Institut de Ciencies de l’Espai (IEEC/CSIC), the Institut de Fisica d’Altes Energies, Lawrence Berkeley National Laboratory, the Ludwig Maximilians Universitat Munchen and the associated Excellence Cluster Universe, the University of Michigan, NSF’s NOIRLab, the University of Nottingham, the Ohio State University, the University of Pennsylvania, the University of Portsmouth, SLAC National Accelerator Laboratory, Stanford University, the University of Sussex, and Texas A\&M University.

BASS is a key project of the Telescope Access Program (TAP), which has been funded by the National Astronomical Observatories of China, the Chinese Academy of Sciences (the Strategic Priority Research Program “The Emergence of Cosmological Structures” Grant \# XDB09000000), and the Special Fund for Astronomy from the Ministry of Finance. The BASS is also supported by the External Cooperation Program of Chinese Academy of Sciences (Grant \# 114A11KYSB20160057), and Chinese National Natural Science Foundation (Grant \# 12120101003, \# 11433005).

The Legacy Survey team makes use of data products from the Near-Earth Object Wide-field Infrared Survey Explorer (NEOWISE), which is a project of the Jet Propulsion Laboratory/California Institute of Technology. NEOWISE is funded by the National Aeronautics and Space Administration.

The Legacy Surveys imaging of the DESI footprint is supported by the Director, Office of Science, Office of High Energy Physics of the U.S. Department of Energy under Contract No. DE-AC02-05CH1123, by the National Energy Research Scientific Computing Center, a DOE Office of Science User Facility under the same contract; and by the U.S. National Science Foundation, Division of Astronomical Sciences under Contract No. AST-0950945 to NOIRLab.

\end{acknowledgements}

\bibliography{virgoha}{}
\bibliographystyle{aasjournal}

\end{document}